\documentclass{elsart}

\usepackage{amssymb}
\usepackage{graphicx}
\usepackage{dcolumn}
\usepackage{bm}
\usepackage{color}
\usepackage{array}
\newcommand{\PreserveBackslash}[1]{\let\temp=\\#1\let\\=\temp}
\newcolumntype{C}[1]{>{\PreserveBackslash\centering}p{#1}}
\newcolumntype{R}[1]{>{\PreserveBackslash\raggedleft}p{#1}}
\newcolumntype{L}[1]{>{\PreserveBackslash\raggedright}p{#1}}

\begin{document}

\begin{frontmatter}

\title{Link Prediction in Complex Networks: A Survey}

\author[1,2]{Linyuan L\"{u}}, \ead{linyuan.lue@unifr.ch}
\author[1,3]{Tao Zhou}\ead{zhutou@ustc.edu}

\address[1]{Web Sciences Center, University of Electronic Science and Technology of China, Chengdu 610054, People's Republic of China}
\address[2]{Department of Physics, University of Fribourg, Chemin du Mus\'{e}e 3, CH-1700 Fribourg, Switzerland}
\address[3]{Department of Modern Physics, University of Science and Technology of China, Hefei 230026, People's Republic of China}

\begin{abstract}

Link prediction in complex networks has attracted increasing
attention from both physical and computer science communities. The
algorithms can be used to extract missing information, identify
spurious interactions, evaluate network evolving mechanisms, and
so on. This article summaries recent progress about link
prediction algorithms, emphasizing on the contributions from
physical perspectives and approaches, such as the
random-walk-based methods and the maximum likelihood methods. We
also introduce three typical applications: reconstruction of
networks, evaluation of network evolving mechanism and
classification of partially labelled networks. Finally, we
introduce some applications and outline future challenges of link
prediction algorithms.

\end{abstract}


\end{frontmatter}

\section{Introduction}

Many social, biological, and information systems can be well
described by networks, where nodes represent individuals, and
links denote the relations or interactions between nodes. The
study of complex networks has therefore become a common focus of
many branches of science. Great efforts have been made to
understand the evolution of networks
\cite{Albert2002,Dorogovtsev2002}, the relations between
topologies and functions \cite{Newman2003,Boccaletti2006}, and the
network characteristics \cite{Costa2007}. An important scientific
issue relevant to network analysis is the so-called
\emph{information retrieval} \cite{Salton1983,Salton1989}, which
aims at finding material of an unstructured nature that satisfies
an information need from large collections \cite{Manning2008}. It
can also be viewed as prediction of relations between words and
documents and is now further extended to stand for a number of
problems on link mining, wherein \emph{link prediction} is the
most fundamental problem that attempts to estimate the likelihood
of the existence of a link between two nodes, based on observed
links and the attributes of nodes \cite{Getoor2005}.

In many biological networks, such as food webs, protein-protein
interaction networks and metabolic networks, whether a link
between two nodes exists must be demonstrated by field and/or
laboratorial experiments, which are usually very costly. Our
knowledge of these networks is very limited, for example, 80\% of
the molecular interactions in cells of Yeast \cite{Yu2008} and
99.7\% of human \cite{Stumpf2008,Amaral2008} are still unknown.
Instead of blindly checking all possible interactions, to predict
based on known interactions and focus on those links most likely
to exist can sharply reduce the experimental costs if the
predictions are accurate enough. Social network analysis also
comes up against the missing data problem
\cite{Schafer2002,Kossinets2006,Neal2008}, where link prediction
algorithms may play a role. In addition, the data in constructing
biological and social networks may contain inaccurate information,
resulting in spurious links \cite{Mering2002,Butts2003}. Link
prediction algorithms can be applied in identifying these spurious
links \cite{Guimera2009}.

Besides helping in analyzing networks with missing data, the link
prediction algorithms can be used to predict the links that may
appear in the future of evolving networks. For example, in online
social networks, very likely but not yet existent links can be
recommended as promising friendships, which can help users in
finding new friends and thus enhance their loyalties to the web
sites. Similar techniques can be applied to evaluate the evolving
mechanism for given networks. For example, many evolving models for
the Internet topology have been proposed: some more accurately
reproduce the degree distribution and the disassortative mixing
pattern \cite{Zhou2004}, some better characterize the $k$-core
structure \cite{Carmi2007}, and so on. Since there are too many
topological features and it is very hard to put weights on them, we
are not easy to judge which model (i.e., which evolving mechanism)
is better than the others. Notice that, each model in principle
corresponds to a link prediction algorithm, and thus we can use the
metrics on prediction accuracy to evaluate the performance of
different models.

Link prediction problem is a long-standing challenge in modern
information science, and a lot of algorithms based on Markov
chains and statistical models have been proposed by computer
science community. However, their works have not caught up the
current progress of the study of complex networks, especially,
they lack serious consideration of the structural characteristics
of networks, like the hierarchical organization
\cite{Sales-Pardo2007} and community structure \cite{Girvan2002},
which may indeed provide useful information and insights for link
prediction. Recently, some physical approaches, such as random
walk processes and maximum likelihood methods, have found
applications in link prediction. This article will give detailed
discussion on these new development.

This article is organized as follows. In the next section, we will
present the link prediction problem and the standard metrics for
performance evaluation. Our tour of link prediction algorithms
starts with the mainstreaming class of algorithms, the so-called
\emph{similarity-based algorithms} \footnote{The similarity
indices between nodes are also called kernels on graphs in some
literature of computer science community \cite{Taylor2004}}, which
are further classified into three categories according to the
information used by the similarity indices: local indices, global
indices and quasi-local indices. In Section 4 and Section 5, we
introduce the maximum likelihood algorithms and probabilistic
models for link prediction. The applications of link prediction
algorithms are presented in Section 6, including the
reconstruction of networks, the evaluation of network evolving
mechanism and the classification of partially labeled networks.
Finally, we outline some future challenges of link prediction
algorithms.

\section{Problem Description and Evaluation Metrics}
Consider an undirected network $G(V,E)$, where $V$ is the set of
nodes and $E$ is the set of links. Multiple links and
self-connections are not allowed. Denote by $U$ the universal set
containing all $\frac{|V|\cdot(|V|-1)}{2}$ possible links, where
$|V|$ denotes the number of elements in set $V$. Then, the set of
nonexistent links is $U-E$. We assume there are some missing links
(or the links that will appear in the future) in the set $U-E$, and
the task of link prediction is to find out these links.

Generally, we do not know which links are the missing or future
links, otherwise we do not need to do prediction. Therefore, to
test the algorithm's accuracy, the observed links, $E$, is
randomly divided into two parts: the training set, $E^T$, is
treated as known information, while the probe set (i.e.,
validation subset), $E^P$, is used for testing and no information
in this set is allowed to be used for prediction. Clearly,
$E^T\cup E^P=E$ and $E^T\cap E^P={\o}$. The advantage of this
random sub-sampling validation is that the proportion of the
training split is not dependent on the number of iterations. But
with this method some links may never be selected in the probe
set, whereas others may be selected more than once, resulting in
statistical bias. This limitation can be overcame by using the
\emph{$K$-fold cross-validation}, in which the observed links is
randomly partitioned into $K$ subsets. Each time one subset is
selected as probe set, the rest $K-1$ constitute the training set.
The cross-validation process is then repeated $K$ times, with each
of the $K$ subsets used exactly once as the probe set. With this
method, all links are used for both training and validation, and
each link is used for prediction exactly once. Clearly, a larger
$K$ will lead to smaller statistical bias yet require more
computation. Some experimental evidences suggested that the
10-fold cross-validation is a very good tradeoff between cost and
performance \cite{Breiman1992,Kohavi1995}. An extreme case called
\emph{leave-one-out} method (i.e., the $|V|$-flod
cross-validation) will be applied in Section 6.2.

Two standard metrics are used to quantify the accuracy of prediction
algorithms: \emph{area under the receiver operating characteristic
curve} (AUC) \cite{Hanely1982} and \emph{Precision}
\cite{Geisser1993,Herlocker2004}. In principle, a link prediction
algorithm provides an ordered list of all non-observed links (i.e.,
$U-E^T$) or equivalently gives each non-observed link, say $(x,y)\in
U-E^T$, a score $s_{xy}$ to quantify its existence likelihood. The
AUC evaluates the algorithm's performance according to the whole
list while the Precision only focuses on the $L$ links with the top
ranks or the highest scores. A detailed introduction of these two
metrics is as follows.

(i) AUC.--- Provided the rank of all non-observed links, the AUC
value can be interpreted as the probability that a randomly chosen
missing link (i.e., a link in $E^P$) is given a higher score than a
randomly chosen nonexistent link (i.e., a link in $U-E$). In the
algorithmic implementation, we usually calculate the score of each
non-observed link instead of giving the ordered list since the
latter task is more time-consuming. Then, at each time we randomly
pick a missing link and a nonexistent link to compare their scores,
if among $n$ independent comparisons, there are $n'$ times the
missing link having a higher score and $n''$ times they have the
same score, the AUC value is
\begin{equation}
\texttt{AUC}=\frac{n'+0.5n''}{n}.
\end{equation}
If all the scores are generated from an independent and identical
distribution, the AUC value should be about 0.5. Therefore, the
degree to which the value exceeds 0.5 indicates how much better the
algorithm performs than pure chance.

(ii) Precision.--- Given the ranking of the non-observed links, the
Precision is defined as the ratio of relevant items selected to the
number of items selected. That is to say, if we take the top-$L$
links as the predicted ones, among which $L_r$ links are right
(i.e., there are $L_r$ links in the probe set $E^P$), then the
Precision equals $L_{r}/L$. Clearly, higher precision means higher
prediction accuracy.

\begin{figure}
\begin{center}
\includegraphics[width=10cm]{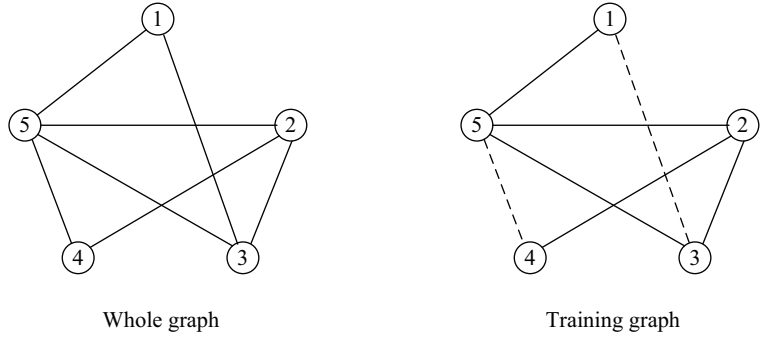}
\caption{An illustration about the calculation of AUC and
Precision.}\label{example}
\end{center}
\end{figure}

Figure ~\ref{example} shows an example of how to calculate the AUC
and Precision. In this simple graph, there are five nodes, seven
existent links and three nonexistent links ($(1,2)$, $(1,4)$ and
$(3,4)$). To test the algorithm's accuracy, we need to select some
existent links as probe links. For instance, we pick $(1,3)$ and
$(4,5)$ as probe links, which are presented by dash lines in the
right plot. Then, any algorithm can only make use of the information
contained in the training graph (presented by solid lines in the
right plot). If an algorithm assigns scores of all non-observed
links as $s_{12}=0.4$, $s_{13}=0.5$, $s_{14}=0.6$, $s_{34}=0.5$ and
$s_{45}=0.6$. To calculate AUC, we need to compare the scores of a
probe link and a nonexistent link. There are in total six pairs:
$s_{13}>s_{12}$, $s_{13}<s_{14}$, $s_{13}=s_{34}$, $s_{45}>s_{12}$,
$s_{45}=s_{14}$ and $s_{45}>s_{34}$. Hence, the AUC value equals
$(3\times1+2\times0.5)/6\approx0.67$. For Precision, if $L=2$, the
predicted links are $(1,4)$ and $(4,5)$. Clearly, the former is
wrong while the latter is right, and thus the Precision equals 0.5.

\section{Similarity-Based Algorithms}

The simplest framework of link prediction methods is the
similarity-based algorithm, where each pair of nodes, $x$ and $y$,
is assigned a score $s_{xy}$, which is directly defined as the
similarity (or called proximity in the literature) between $x$ and
$y$. All non-observed links are ranked according to their scores,
and the links connecting more similar nodes are supposed to be of
higher existence likelihoods. In despite of its simplicity, the
study on similarity-based algorithms is the mainstream issue. In
fact, the definition of node similarity is a nontrivial challenge.
Similarity index can be very simple or very complicated and it may
work well for some networks while fails for some others. In
addition, the similarities can be used in a more skilled way, such
as being locally integrated under the \emph{collaborative
filtering}\footnote{Collaborative filtering is the process of
filtering for information or patterns using techniques involving
collaboration among multiple agents, viewpoints, data sources, etc.
\cite{Su2009}} framework \cite{Huang2005}.

Node similarity can be defined by using the essential attributes of
nodes: two nodes are considered to be similar if they have many
common features \cite{Lin1998}. However, the attributes of nodes are
generally hidden, and thus we focus on another group of similarity
indices, called \emph{structural similarity}, which is based solely
on the network structure. The structural similarity indices can be
classified in various ways, such as local vs. global, parameter-free
vs. parameter-dependent, node-dependent vs. path-dependent, and so
on. The similarity indices can also be sophisticatedly classified as
\emph{structural equivalence} and \emph{regular equivalence}. The
former embodies a latent assumption that the link itself indicated a
similarity between two endpoints (see, for example, the
\emph{Leicht-Holme-Newman index} \cite{Leicht2006} and
\emph{transferring similarity} \cite{Sun2009}), while the latter
assumes that two nodes are similar if their neighbors are similar.
Readers are encouraged to see Ref. \cite{White1983} for the
mathematical definition of regular equivalence and Ref.
\cite{Holme2005} for a recent application on the prediction of
protein functions.

Here we adopt the simplest method, where 20 similarity indices are
classified into three categories: the former 10 are local indices,
followed by 7 global indices, and the last 3 are quasi-local
indices, which do not require global topological information but
make use of more information than local indices.

\subsection{Local Similarity Indices}

(1) \emph{Common Neighbours} (CN). For a node $x$, let $\Gamma(x)$
denote the set of neighbors of $x$. In common sense, two nodes, $x$
and $y$, are more likely to have a link if they have many common
neighbors. The simplest measure of this neighborhood overlap is the
directed count, namely
\begin{equation}
s^{CN}_{xy}=|\Gamma(x)\cap \Gamma(y)|. \label{Sim1}
\end{equation}
where $|Q|$ is the cardinality of the set $Q$. It is obvious that
$s_{xy}=(A^2)_{xy}$, where $A$ is the adjacency matrix:
$A_{xy}=1$ if $x$ and $y$ are directly connected and $A_{xy}=0$
otherwise. Note that, $(A^2)_{xy}$ is also the number of different
paths with length 2 connecting $x$ and $y$. Newman
\cite{Newman2001a} used this quantity in the study of collaboration
networks, showing a positive correlation between the number of
common neighbors and the probability that two scientists will
collaborate in the future. Kossinets and Watts \cite{Kossinets2006}
analyzed a large-scale social network, suggesting that two students
having many mutual friends are very probable to be friend in the
future.

(2) \emph{Salton Index} \cite{Salton1983}. It is defined as
\begin{equation}
s^{Salton}_{xy}=\frac{|\Gamma(x)\cap \Gamma(y)|}{\sqrt{k_x\times
k_y}}, \label{Sim2}
\end{equation}
where $k_x$ is the degree of node $x$. The Salton index is also
called the cosine similarity in the literature.

(3) \emph{Jaccard Index} \cite{Jaccard1901}. This index was proposed
by Jaccard over a hundred years ago, and is defined as:
\begin{equation}
s^{Jaccard}_{xy}=\frac{|\Gamma(x)\cap \Gamma(y)|}{|\Gamma(x)\cup
\Gamma(y)|}. \label{Sim3}
\end{equation}

(4) \emph{S{\o}rensen Index} \cite{Sorensen1948}. This index is used
mainly for ecological community data, and is defined as
\begin{equation}
s^{S{\o}rensen}_{xy}=\frac{2|\Gamma(x)\cap \Gamma(y)|}{k_x+k_y}.
\label{Sim4}
\end{equation}

(5) \emph{Hub Promoted Index} (HPI) \cite{Ravasz2002}. This index is
proposed for quantifying the topological overlap of pairs of
substrates in metabolic networks, and is defined as
\begin{equation}
s^{HPI}_{xy}=\frac{|\Gamma(x)\cap \Gamma(y)|}{\min\{k_x,k_y\}}.
\label{Sim5}
\end{equation}
Under this measurement, the links adjacent to hubs are likely to be
assigned high scores since the denominator is determined by the
lower degree only.

(6) \emph{Hub Depressed Index} (HDI). Analogously to the above
index, we also consider a measurement with the opposite effect on
hubs, defined as
\begin{equation}
s^{HDI}_{xy}=\frac{|\Gamma(x)\cap \Gamma(y)|}{\max\{k_x,k_y\}}.
\label{Sim6}
\end{equation}

(7) \emph{Leicht-Holme-Newman Index} (LHN1) \cite{Leicht2006}. This
index assigns high similarity to node pairs that have many common
neighbors compared not to the possible maximum, but to the expected
number of such neighbors. It is defined as
\begin{equation}
s^{LHN1}_{xy}=\frac{|\Gamma(x)\cap \Gamma(y)|}{k_x\times k_y},
\label{Sim7}
\end{equation}
where the denominator, $k_x\times k_y$, is proportional to the
expected number of common neighbors of nodes $x$ and $y$ in the
configuration model \cite{Molloy1995}. We use the abbreviation LHN1
to distinguish this index to another index (named as LHN2 index)
also proposed by Leicht, Holme and Newman.

(8) \emph{Preferential Attachment Index} (PA). The mechanism of
preferential attachment can be used to generate evolving scale-free
networks, where the probability that a new link is connected to the
node $x$ is proportional to $k_x$ \cite{Barabasi1999}. A similar
mechanism can also lead to scale-free networks without growth
\cite{Xie2008}, where at each time step, an old link is removed and
a new link is generated. The probability that this new link will
connect $x$ and $y$ is proportional to $k_x\times k_y$. Motivated by
this mechanism, the corresponding similarity index can be defined as
\begin{equation}
s^{PA}_{xy}=k_x\times k_y, \label{Sim8}
\end{equation}
which has been widely used to quantify the functional significance
of links subject to various network-based dynamics, such as
percolation \cite{Holme2002}, synchronization \cite{Yin2006} and
transportation \cite{Zhang2007}. Note that this index does not
require the information of the neighborhood of each node, as a
consequence, it has the least computational complexity.

(9) \emph{Adamic-Adar Index} (AA) \cite{Adamic2003}. This index
refines the simple counting of common neighbors by assigning the
less-connected neighbors more weight, and is defined as
\begin{equation}
s^{AA}_{xy}=\sum_{z\in \Gamma(x)\cap \Gamma(y)}\frac{1}{\log k_z}.
\label{Sim9}
\end{equation}

(10) \emph{Resource Allocation Index} (RA) \cite{Zhou2009}. This
index is motivated by the resource allocation dynamics on complex
networks \cite{Ou2007}. Consider a pair of nodes, $x$ and $y$, which
are not directly connected. The node $x$ can send some resource to
$y$, with their common neighbors playing the role of transmitters.
In the simplest case, we assume that each transmitter has a unit of
resource, and will equally distribute it to all its neighbors.
The similarity between $x$ and $y$ can be defined as the amount of
resource $y$ received from $x$, which is:
\begin{equation}
s^{RA}_{xy}=\sum_{z\in \Gamma(x)\cap \Gamma(y)}\frac{1}{k_z}.
\label{Sim10}
\end{equation}
Clearly, this measure is symmetric, namely $s_{xy}=s_{yx}$. Note
that, although resulting from different motivations, the AA index
and RA index have very similar form. Indeed, they both depress the
contribution of the high-degree common neighbors. AA index takes the
form $(\log k_z)^{-1}$ while RA index takes the form $k_z^{-1}$. The
difference is insignificant when the degree, $k_z$, is small, while
it is considerable when $k_z$ is large. In a word, RA index punishes
the high-degree common neighbors more heavily than AA.

\begin{table}
\caption{Accuracies of different local similarity indices subject to
link prediction, measured by the AUC value. Each number is obtained
by averaging over 10 implementations with independently random
partitions of testing set (90\%) and probe set (10\%). The entries
corresponding to the highest accuracies among these 10 indices are
emphasized in black. The six real networks for testing are a
protein-protein interaction network (PPI) \cite{Mering2002}, a
co-authorship network of scientists who are themselves publishing on
the topic of network science (NS) \cite{Newman2006}, an electrical
power-grid of the western US (Grid) \cite{Watts1998}, a network of
the US political blogs (PB) \cite{Ackland2005}, a router-level
Internet collected by \emph{Rocketfuel Project} (INT)
\cite{Spring2004}, and a network of the US air transportation system
(USAir) \cite{Batageli}. Detailed information about these networks
can be found in Ref. \cite{Zhou2009}.}
\begin{center}
\begin{tabular} {ccccccc}
  \hline \hline
   Indices     & PPI  &  NS  &  Grid & PB & INT & USAir \\
   \hline
CN     & 0.889  &  \textbf{0.933}  &  \textbf{0.590} & 0.925 & \textbf{0.559} & 0.937 \\
Salton     & 0.869  &  0.911  &  0.585 & 0.874 & 0.552 & 0.898  \\
Jaccard     & 0.888  &  \textbf{0.933}  &  \textbf{0.590} & 0.882 & \textbf{0.559 }& 0.901  \\
S{\o}rensen     & 0.888  &  \textbf{0.933}  &  \textbf{0.590} & 0.881 & \textbf{0.559} & 0.902  \\
HPI     & 0.868  &  0.911  &  0.585 & 0.852 & 0.552 & 0.857  \\
HDI     & 0.888  &  \textbf{0.933}  & \textbf{0.590}  & 0.877 & \textbf{0.559} & 0.895  \\
LHN1     & 0.866  &  0.911  &  0.585 & 0.772 & 0.552 & 0.758  \\
PA     & 0.828  &  0.623  &  0.446 & 0.907 & 0.464 & 0.886  \\
AA     & 0.888  &  0.932  &  \textbf{0.590} & 0.922 & \textbf{0.559} & 0.925 \\
RA     & \textbf{0.890}  &  \textbf{0.933}  &  \textbf{0.590} & \textbf{0.931} & \textbf{0.559} & \textbf{0.955} \\
   \hline \hline
\end{tabular}
\end{center}
\label{Local indices}
\end{table}

Liben-Nowell \emph{et al.} \cite{Liben-Nowell2007} and Zhou \emph{et
al.} \cite{Zhou2009} systematically compared a number of local
similarity indices on many real networks: the former
\cite{Liben-Nowell2007} focuses on social collaboration networks and
the latter \cite{Zhou2009} considers disparate networks including
the protein-protein interaction network, electronic grid, Internet,
US airport network, etc. According to extensive experimental results
on real networks (see results in Table \ref{Local indices}), the RA
index performs best, while AA and CN indices have the second best
overall performance among all the above-mentioned local indices.

The PA index has the worst overall performance, yet we are
interested in it for it requires the least information. Notice that,
PA performs even worst than pure chance for the Internet at router
level and the power grid. In these two networks, the nodes have
well-defined positions and the links are physical lines. Actually,
geography plays a significant role and links with very long
geographical distances are rare. As local centers, the high-degree
nodes have longer geographical distances to each other than average,
and thus have a lower probability of directly connecting to each
other, which leads to the bad performance of PA. In contrast,
although USAir has well-defined geographical positions of nodes, its
links are not physical. Empirical data has demonstrated that the
number of airline flights is not very sensitive to the geographical
distance within a big range \cite{Gastner2006,Liu2007} (another
topological evidence for the relatively good performance of PA on
USAir is the so-called rich-club phenomenon
\cite{Zhou2004b,Colizza2006}). The LHN1 index performs the second
worst, however, compared with all other neighborhood-based indices,
it is very good at uncovering the missing links connecting two
small-degree nodes \cite{Zhu2010}.

Recently, Pan \emph{et al.} \cite{Pan2010} have compared all the
local indices appeared in Ref. \cite{Zhou2009} in a similarity-based
community detection algorithm, and their experimental results again
indicate that the RA index performs best. Wang \emph{et al.}
\cite{Wang2009} have applied the RA index to estimate the weights
between stations in Chinese railway, which shows better performance
than the CN index. In addition, the RA index for bipartite networks
can be applied in personalized recommendation with higher accuracy
than the classical collaborative filtering \cite{Zhou2007}.

\subsection{Global Similarity Indices}

(11) \emph{Katz Index} \cite{Katz1953}. This index is based on the
ensemble of all paths, which directly sums over the collection of
paths and is exponentially damped by length to give the shorter
paths more weights. The mathematical expression reads
\begin{equation}
s^{Katz}_{xy}=\sum^{\infty}_{l=1}\beta^l\cdot|paths^{<l>}_{xy}|=\beta
A_{xy}+\beta^2(A^2)_{xy}+\beta^3(A^3)_{xy}+\cdots, \label{Sim11}
\end{equation}
where $paths^{<l>}_{xy}$ is the set of all paths with length $l$
connecting $x$ and $y$, and $\beta$ is a free parameter (i.e., the
damping factor) controlling the path weights. Obviously, a very
small $\beta$ yields a measurement close to CN, because the long
paths contribute very little. The similarity matrix can be written
as
\begin{equation}
S^{Katz}=(I-\beta A)^{-1}-I. \label{Sim11-matrix}
\end{equation}
Note that, $\beta$ must be lower than the reciprocal of the largest
eigenvalue of matrix $A$ to ensure the convergence of Eq.
\ref{Sim11}.

(12) \emph{Leicht-Holme-Newman Index} (LHN2) \cite{Leicht2006}. This
index is a variant of the Katz index. Based on the concept that two
nodes are similar if their immediate neighbors are themselves
similar, one obtains a self-consistent matrix formulation
\begin{equation}
S=\phi{AS}+\psi{I}=\psi(I-\phi{A})^{-1}=\psi(I+\phi{A}+\phi^2A^2+\cdots),
\label{Sim12-1}
\end{equation}
where $\phi$ and $\psi$ are free parameters controlling the balance
between the two components of the similarity. Setting $\psi=1$, it
is very similar to the Katz index. Note that $(A^l)_{xy}$ is equal
to the number of paths of length $l$ from $x$ to $y$. The expected
value of $(A^l)_{xy}$, namely $E[(A^l)_{xy}]$, equals
$(k_xk_y/2M)\lambda_1^{l-1}$, where $\lambda_1$ is the largest
eigenvalue of $A$ and $M$ is the total number of edges in the
network. Replace $(A^l)_{xy}$ in Eq.~\ref{Sim12-1} with
$(A^l)_{xy}/E[(A^l)_{xy}]$, we obtain the expression:
\begin{equation}
s^{LHN2}_{xy}=\delta_{xy}+\frac{2M}{k_xk_y}\sum_{l=0}^{\infty}\phi^l\lambda^{1-l}(A^l)_{xy}
=\left[1-\frac{2M\lambda_1}{k_xk_y}\right]\delta_{xy}+\frac{2M\lambda_1}{k_xk_y}\left[(I-\frac{\phi}{\lambda_1}A)^{-1}\right]_{xy},
\label{Sim12-2}
\end{equation}
where $\delta_{xy}$ is the \emph{Kronecker's function}. Since the
first item is a diagonal matrix, it can be dropped and thus we
arrive to a compact expression
\begin{equation}
S=2m{\lambda_1}D^{-1}(I-\frac{\phi{A}}{\lambda_1})^{-1}D^{-1},
\label{Sim12}
\end{equation}
where $D$ is the degree matrix with $D_{xy}=\delta_{xy}k_{x}$ and
$\phi$ $(0<\phi<1)$ is a free parameter. The choosing of $\phi$
depends on the investigated network, and smaller $\phi$ assigns more
weights on shorter paths.

(13) \emph{Average Commute Time} (ACT). Denote by $m(x,y)$ the
average number of steps required by a random walker starting from
node $x$ to reach node $y$, the average commute time between $x$ and
$y$ is
\begin{equation}
n(x,y)=m(x,y)+m(y,x), \label{ACT-random-walk-expression}
\end{equation}
which can be obtained in terms of the pseudoinverse of the Laplacian
matrix, $L^{+}$ ($L=D-A$), as \cite{Klein1993,Fouss2007}:
\begin{equation}
n(x,y)=M(l_{xx}^{+}+l_{yy}^{+}-2l_{xy}^{+}), \label{ACT}
\end{equation}
where $l_{xy}^{+}$ denotes the corresponding entry in $L^{+}$.
Assuming two nodes are more similar if they have a smaller average
commute time, then the similarity between the nodes $x$ and $y$ can
be defined as the reciprocal of $n(x,y)$, namely (the constant
factor $M$ is removed)
\begin{equation}
s^{ACT}_{xy}=\frac{1}{l_{xx}^{+}+l_{yy}^{+}-2l_{xy}^{+}}.
\label{Sim13}
\end{equation}

(14) \emph{Cosine based on $L^+$}. This index is an
inner-product-based measure. In the Euclidean space spanned by
$v_x=\Lambda^{\frac{1}{2}}U^T\vec{e_x}$, where $U$ is an orthonormal
matrix made of the eigenvectors of $L^+$ ordered in decreasing order
of corresponding eigenvalue $\lambda_x$,
$\Lambda=\mathrm{diag}(\lambda_x)$, $\vec{e_x}$ is an $N\times 1$
vector with the $x$th element equal to $1$ and others all equal to
$0$, and $T$ is the matrix transposition, the pseudoinverse of the
Laplacian matrix are the inner products of the node vectors,
$l_{xy}^+=v_x^Tv_y$. Accordingly, the cosine similarity is defined
as the cosine of the node vectors, namely \cite{Fouss2007}:
\begin{equation}
s^{cos^+}_{xy}=cos(x,y)^+=\frac{v_x^Tv_y}{|v_x|\cdot|v_y|}=\frac{l_{xy}^+}{\sqrt{l_{xx}^+
\cdot l_{yy}^+}}.
\end{equation}

(15) \emph{Random Walk with Restart} (RWR). This index is a direct
application of the PageRank algorithm \cite{Brin1998}. Consider a
random walker starting from node $x$, who will iteratively moves to
a random neighbor with probability $c$ and return to node $x$ with
probability $1-c$. Denote by $q_{xy}$ the probability this random
walker locates at node $y$ in the steady state, we have
\begin{equation}
\vec{q_x}= c P^{T} \vec{q_x}+(1-c)\vec{e_x}, \label{RWR1}
\end{equation}
where $P$ is the transition matrix with $P_{xy}=1/k_{x}$ if $x$ and
$y$ are connected, and $P_{xy}=0$ otherwise. The solution is
straightforward, as
\begin{equation}
\vec{q_x} = (1-c) (I-cP^{T})^{-1}\vec{e_x}. \label{RWR2}
\end{equation}
The RWR index is thus defined as
\begin{equation}
s^{RWR}_{xy}=q_{xy}+q_{yx}. \label{Sim15}
\end{equation}
A fast algorithm to calculate this index was proposed by Tong
\emph{et al.} \cite{Tong2006}, and the application of this index to
recommender systems can be found in Ref. \cite{Shang2009}.

(16) \emph{SimRank} \cite{Jeh2002}. Similar to the LHN2, SimRank is
defined in a self-consistent way, according to the assumption that
two nodes are similar if they are connected to similar nodes:
\begin{equation}
s^{SimRank}_{xy}=C\cdot\frac{\sum_{z\in{\Gamma(x)}}\sum_{z'\in{\Gamma(y)}}s^{SimRank}_{zz'}}{k_x\cdot
k_y} \label{Sim16}
\end{equation}
where $s_{xx}=1$ and $C\in{[0,1]}$ is the decay factor. The SimRank
can also be interpreted by the random-walk process, that is,
$s^{SimRank}_{xy}$ measures how soon two random walkers,
respectively starting from nodes $x$ and $y$, are expected to meet
at a certain node.

(17) \emph{Matrix Forest Index} (MFI) \cite{Chebotarev1997}. This
index is defined as
\begin{equation}
S=(I+L)^{-1}, \label{Sim17}
\end{equation}
where the similarity between $x$ and $y$ can be understood as the
ratio of the number of spanning rooted forests such that nodes $x$
and $y$ belong to the same tree rooted at $x$ to all spanning rooted
forests of the network (see details in Ref. \cite{Chebotarev1997}).
A parameter-dependent variant of MFI is
\begin{equation}
S=(I+\alpha{L})^{-1},\texttt{      }\alpha>0. \label{variant MFI}
\end{equation}
This index has been applied to quantify the similarity between
nodes on collaborative recommendation task \cite{Fouss2006}. The
results indicate that a simple nearest-neighbors rule based on
similarity measured by MFI performs best.

Comparing with the local similarity indices, the global ones ask for
the whole topological information. Although the global indices can
provide much more accurate prediction than the local ones, they
suffer two big disadvantages: (i) the calculation of a global index
is very time-consuming, and is usually infeasible for large-scale
networks; (ii) sometimes, the global topological information is not
available, especially if we would like to implement the algorithm in
a decentralized manner. As we will show in the next subsection, a
promising tradeoff is the quasi-local indices, which considers more
information than local indices while abandons the superfluous
information that makes no contribution or very little contribution
to the prediction accuracy.

\subsection{Quasi-Local Indices}

(18) \emph{Local Path Index} (LP) \cite{Zhou2009,Lu2009}. To provide
a good tradeoff of accuracy and computational complexity, we here
introduce an index that takes consideration of local paths, with
wider horizon than CN. It is defined as
\begin{equation}
S^{LP}=A^2+\epsilon A^3,
\end{equation}
where $\epsilon$ is a free parameter. Clearly, this measure
degenerates to CN when $\epsilon=0$. And if $x$ and $y$ are not
directly connected (this is the case we are interested in),
$(A^3)_{xy}$ is equal to the number of different paths with length 3
connecting $x$ and $y$. This index can be extended to account for
higher-order paths, as
\begin{equation}
S^{LP(n)}=A^2+\epsilon A^3+\epsilon^2 A^4+\cdots+\epsilon^{n-2} A^n,
\end{equation}
where $n>2$ is the maximal order. With the increasing of $n$, this
index asks for more information and computation. Especially, when
$n\rightarrow\infty$, $S^{LP(n)}$ will be equivalent to the Katz
index that takes into account all paths in the network. The
computational complexity of this index in an uncorrelated network is
$\mathbb{O}(N\langle k\rangle ^n)$, which grows fast with the
increasing of $n$ and will exceed the complexity for calculating the
Katz index (approximate to $\mathbb{O}(N^3)$) for large $n$.
Experimental results show that the optimal $n$ is positively
correlated with the average shortest distance of the network
\cite{Lu2009}.

The LP index performs remarkably better than the neighborhood-based
indices, such as RA, AA and CN \cite{Zhou2009}. It is because the
neighborhood information is less distinguishable and two node pairs
are of high probability to be assigned the same similarity scores.
Taking INT as an example, there are more than $10^7$ node pairs,
99.59\% of which are assigned zero score by CN. For all the node
pairs having scores higher than 0, 91.11\% are assigned score 1, and
4.48\% are assigned score 2. Using a little bit more information
involving the next nearest neighbors may break the ``degeneracy of
the states¡± and make the similarity scores more distinguishable.
This is the reason why the LP index largely improves the prediction
accuracy.

The comparison of LP index with other two path-dependent global
indices, the Katz and LHN2 indices, is shown in Table ~\ref{Local
Path}. Overall speaking, the Katz index performs best subject to the
AUC value, while the LP index is the best for the Precision. For the
network with small average shortest distance (e.g., USAir and PB),
LP index gives the most accurate predictions for both AUC and
Precision. In a word, the LP index provides competitively good
predictions while asks for much lighter computation compared with
the global indices.

\begin{table}
\caption{Accuracies of the three path-dependent similarity indices,
measured by AUC and precision. Here only the main components of
example networks are considered (see Ref. \cite{Lu2009} for detailed
information). Each number is obtained by averaging over 10
independent realizations. The entries corresponding to the highest
accuracies are emphasized in black. For LP, Katz and LHN2 indices,
the AUC values are corresponding to the optimal parameter which will
be used to calculate their corresponding precision where we set
$L=100$. For USAir, the optimal value of $\epsilon$ is negative (see
the explanation in Ref.~\cite{Zhou2009}). LP* denotes the LP index
with a fixed parameter $\epsilon=0.01$ (for USAir $\epsilon=-0.01$).
The very small difference between the optimal case and the case with
$\epsilon=0.01$ suggests that in the real application, one can
directly set $\epsilon$ as a very small number, instead of finding
out its optimum that may cost much time. This again supports our
motivation that the essential advantage of the uasge of the second
order neighborhood is to improve the distinguishability of the
similarity scores.}
\begin{center}
\begin{tabular} {ccccccc}
  \hline \hline
   AUC     & PPI  &  NS  &  Grid & PB & INT & USAir \\
   \hline
   LP & 0.970 & \textbf{0.988} & 0.697 & \textbf{0.941} & 0.943 & \textbf{0.960}\\
   LP* & 0.970 & \textbf{0.988} & 0.697 & 0.939 & 0.941 & 0.959 \\
   Katz & \textbf{0.972} & \textbf{0.988} & \textbf{0.952} & 0.936 & \textbf{0.975} & 0.956 \\
   LHN2 & 0.968 & 0.986 & 0.947 & 0.769 & 0.959 & 0.778 \\
   \hline \hline
   Precision     & PPI  &  NS  &  Grid & PB & INT & USAir \\
   \hline
   LP & \textbf{0.734} & \textbf{0.292} & \textbf{0.132} & \textbf{0.519} & \textbf{0.557} & \textbf{0.627}\\
   LP* & \textbf{0.734}  & \textbf{0.292}  & \textbf{0.132}  &0.469  &0.121  &\textbf{0.627} \\
   Katz & 0.719 & 0.290 & 0.063 & 0.456 & 0.368 & 0.623  \\
   LHN2 &0 &0.060& 0.005 &0&0&0.005\\
   \hline \hline
    \end{tabular}\label{Local Path}
\end{center}
\end{table}

(19) \emph{Local Random Walk} (LRW) \cite{Liu2010}. To measure the
similarity between nodes $x$ and $y$, a random walker is initially
put on node $x$ and thus the initial density vector
$\vec{\pi}_{x}(0)=\vec{e_x}$. This density vector evolves as
$\vec{\pi}_x(t+1)=P^T\vec{\pi}_x(t)$ for $t\geq 0$. The LRW index at
time step $t$ is thus defined as
\begin{equation}
s^{LRW}_{xy}(t)=q_x\pi_{xy}(t)+q_y\pi_{yx}(t). \label{Sim19}
\end{equation}
where $q$ is the initial configuration function. In Ref.
\cite{Liu2010} Liu and L\"u applied a simple form determined by node
degree, namely $q_x=\frac{k_x}{M}$.

(20) \emph{Superposed Random Walk} (SRW) \cite{Liu2010}. Similar to
the RWR index, Liu and L\"u \cite{Liu2010} proposed the SRW index,
where the random walker is continuously released at the starting
point, resulting in a higher similarity between the target node and
the nodes nearby. The mathematical expression reads
\begin{equation}
s^{SRW}_{xy}(t)=\sum_{\tau=1}^{t}s_{xy}^{LRW}(\tau)=\sum_{\tau=1}^{t}{[q_x\pi_{xy}(\tau)+q_y\pi_{yx}(\tau)]},
\label{Sim20}
\end{equation}
where $t$ denotes the time steps.

Liu and L\"u \cite{Liu2010} systematically compared these two
indices, LRW and SRW, with five other indices, including three local
(or quasi-local) indices, CN, RA and LP, and two other
random-walk-based global indices, ACT and RWR, as well as the
\emph{hierarchical structure method} (HSM) proposed by Clauset,
Moore and Newman \cite{Clauset2008} (see Section 4.1 for the
detailed introduction of HSM). According to the experimental results
(see Table 3), LRW and SRW methods perform better than other indices
with their respective optimal walking step positively correlated
with the average shortest distance of the network.

\begin{table}
\caption{Comparison of algorithms' accuracy quantified by AUC and
Precision. For each network, the training set contains 90\% of the
known links. Each number is obtained by averaging over 1000
implementations with independently random divisions of training set
and probe set. The parameters $\varepsilon=10^{-3}$ for LP (for
USAir, $\varepsilon=-10^{-3}$) and $c=0.9$ for RWR.  The numbers
inside the brackets denote the optimal step of LRW and SRW indices.
For example, 0.972(2) means the optimal AUC is obtained at the
second step of LRW. The highest accuracy in each line is emphasized
in black. For HSM, 5000 samples of dendrograms for each
implementation are generated.}
\begin{center}
\small \begin{tabular}{ccccccccc}
  \hline \hline
   \textbf{AUC}    &CN &RA &LP  &ACT &RWR &HSM &LRW &SRW  \\
   \hline
   USAir   &0.954&0.972  &0.952   &0.901  &0.977   & 0.904 &0.972(2)  &\textbf{0.978}(3) \\
   NetScience  &0.978  &0.983   &0.986  &0.934   &\textbf{0.993}  &0.930  &0.989(4) &0.992(3)\\
   Power  &0.626  & 0.626   & 0.697 &0.895   &0.760  &0.503  & 0.953(16)& \textbf{0.963}(16) \\
   Yeast  &0.915  &0.916   &0.970  &0.900   &0.978  & 0.672 &0.974(7)&\textbf{0.980}(8)\\
   C.elegans  &0.849  &0.871   &0.867  &0.747   &0.889  & 0.808 &0.899(3)&\textbf{0.906}(3)\\
   \hline \hline
   \textbf{Precision}    &CN &RA &LP  &ACT &RWR &HSM &LRW &SRW \\
   \hline
   USAir   &0.59  &0.64   &0.61  &0.49   &0.65  &0.28  &0.64(3)&\textbf{0.67}(3)\\
   NetScience  & 0.26 &0.54  &0.30 &0.19   &\textbf{0.55}  &0.25  &0.54(2)&0.54(2)\\
   Power  &0.11  &0.08   &\textbf{0.13}  & 0.08  & 0.09 &0.00  &0.08(2)&0.11(3)\\
   Yeast  &0.67  &0.49   &0.68  &0.57   &0.52  &0.84  &\textbf{0.86}(3)&0.73(9) \\
   C.elegans &0.12  &0.13   &\textbf{0.14}  &0.07   &0.13  & 0.08 &\textbf{0.14}(3) &\textbf{0.14}(3) \\
   \hline \hline
\end{tabular}
\end{center}
\label{compare}
\end{table}

Furthermore, the computational complexity of LRW and SRW is lower
than ACT and RWR whose time complexity in calculating inverse and
pseudoinverse is approximately $\mathbb{O}(N^3)$, while the time
complexity of $n$-steps LRW and SRW are approximately
$\mathbb{O}(N\langle k\rangle^n)$, ignoring degree heterogeneity of
the network. That is to say, when $n$ is small LRW and SRW run much
faster than other random-walk-based global similarity indices. The
advantage of LRW and SRW for their low calculation complexity is
prominent especially in the huge size (i.e. large $N$) and sparse
(i.e. small $\langle k\rangle$) networks. For example, LRW or SRW
for power grid is thousands time faster than ACT, $cos^+$ and RWR,
even for $n\simeq 10$ \cite{Liu2010}.

With the similar motivation of LRW and SRW, Mantrach \emph{et al.}
recently proposed a bounded normalized random walk with restart
algorithm (see Eq.~\ref{RWR1} for the definition of RWR), and
applied it to address the classification problem
\cite{Mantrach2010}. With this method both complexities of time
and space can be reduced.

\section{Maximum Likelihood Methods}
This section will introduce two recently proposed algorithms based
on the maximum likelihood estimation. These algorithms presuppose
some organizing principles of the network structure, with the
detailed rules and specific parameters obtained by maximizing the
likelihood of the observed structure. Then, the likelihood of any
non-observed link can be calculated according to those rules and
parameters.

From the viewpoint of practical applications, an obvious drawback of
the maximum likelihood methods is that it is very time consuming. A
well designed algorithm is able to handle networks with up to a few
thousand nodes in a reasonable time, but will definitely fail to
deal with the huge online networks that often consist of millions of
nodes. In addition, the maximum likelihood methods are probably not
among the most accurate ones (see, for example, the comparison
between hierarchical structure model and some typical
similarity-based methods in Table \ref{compare}). However, the
maximum likelihood methods provide very valuable insights into the
network organization, which can not be gained from the
similarity-based algorithms or the probabilistic models.

\subsection{Hierarchical Structure Model}
Empirical evidence indicates that many real networks are
hierarchically organized, where nodes can be divided into groups,
further subdivided into groups of groups, and so forth over multiple
scales \cite{Sales-Pardo2007} (e.g., metabolic networks
\cite{Ravasz2002} and brain networks \cite{Zhou2006}). As Redner
said \cite{Redner2008}, focusing on the hierarchical structure
inherent in social and biological networks might provide a smart way
to find missing links. Clauset, Moore and Newman \cite{Clauset2008}
proposed a general technique to infer the hierarchical organization
from network data and further applied it to predict the missing
links.

\begin{figure}
\centering
\includegraphics[scale=0.6]{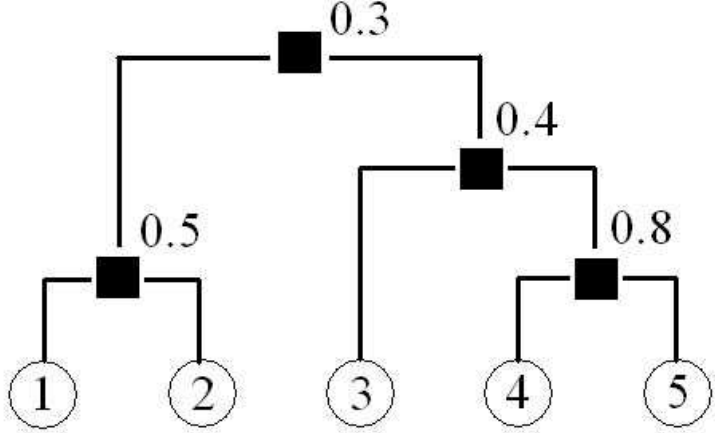}
\caption{Illustration of a dendrogram of a network with 5 nodes.
Accordingly, the connecting probability of nodes 1 and 2 is 0.5, of
nodes 1 and 3 is 0.3, of nodes 3 and 4 is 0.4.}
\label{hierarchical-organization}
\end{figure}

\begin{figure}
\centering
\includegraphics[scale=0.4]{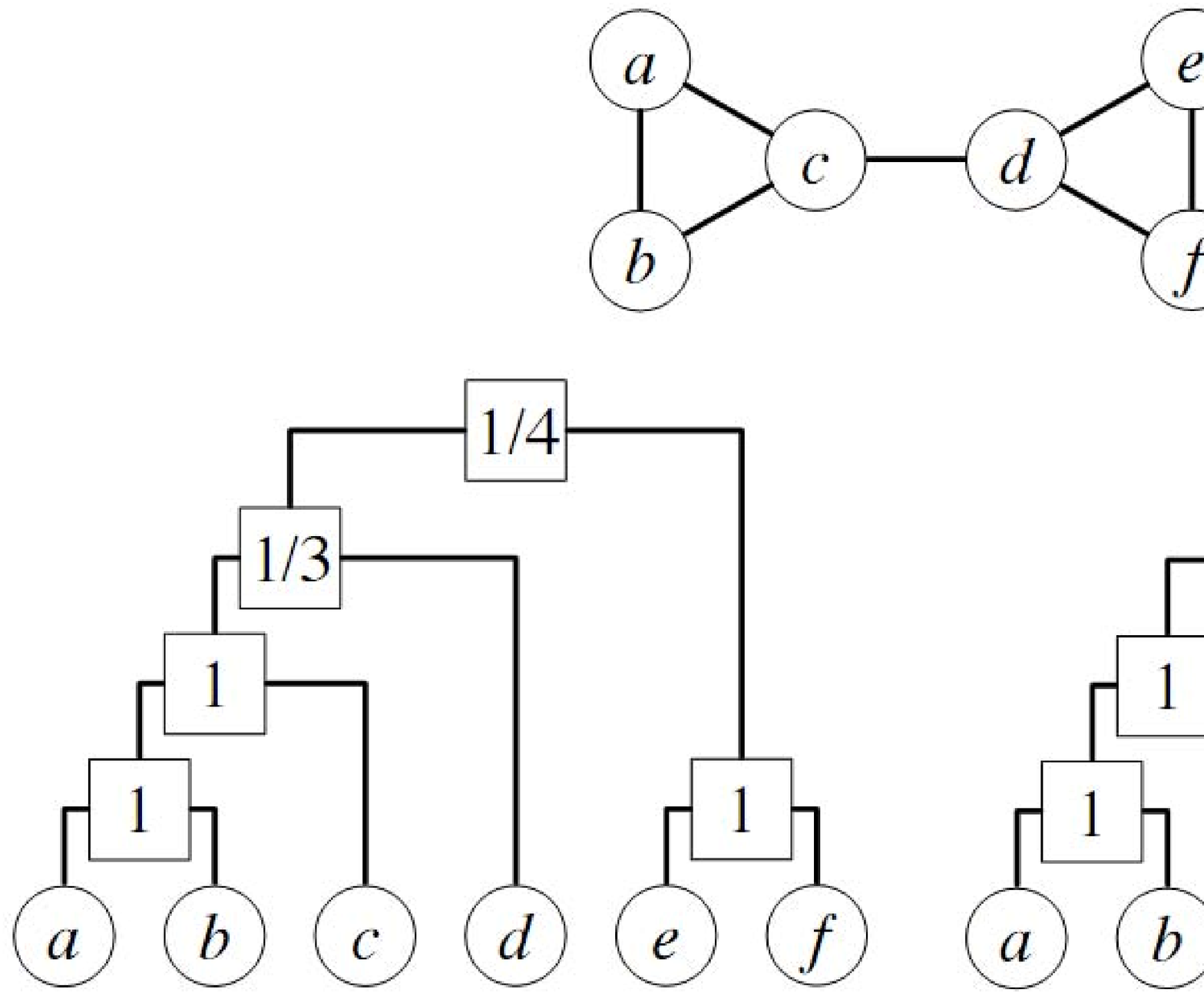}
\caption{The likelihood of two possible dendrograms for an example
network consisting of 6 nodes. The interval nodes are labeled with
the maximum likelihood probability obtained by Eq. \ref{optimal
p_r}. The likelihoods are $\mathcal{L}(D_1)\approx 0.00165$ (left
dendrogram) and $\mathcal{L}(D_1)\approx 0.0433$ (right dendrogram).
Reprinted figure with permission from \cite{Clauset2008}, copyright
is held by Nature Publishing Group.} \label{hierarchical-inferring}
\end{figure}

As shown in Fig. \ref{hierarchical-organization}, the hierarchical
structure of a network can be represented by a dendrogram with $N$
leaves (corresponding to the nodes of the network) and $N-1$
internal nodes. Clauset, Moore and Newman \cite{Clauset2008}
introduced a simple model where each internal node $r$ is associated
with a probability $p_r$ and the connecting probability of a pair of
nodes (leaves) is equal to $p_{r'}$ where $r'$ is the lowest common
ancestor of these two nodes. Given a real network $G$ and a
dendrogram $D$, let $E_r$ be the number of edges in $G$ whose
endpoints have $r$ as their lowest common ancestor in $D$, and let
$L_r$ and $R_r$, respectively, be the number of leaves in the left
and right subtrees rooted at $r$. Then the likelihood of the
dendrogram $D$ together with a set of $p_r$ is
\begin{equation}
\mathcal{L}(D,\{p_r\})=\prod_rp_r^{E_r}(1-p_r)^{L_rR_r-E_r}.
\label{Likelihood hierarchical structures}
\end{equation}
For a fixed $D$, it is obvious that
\begin{equation}
p_r^*=\frac{E_r}{L_rR_r} \label{optimal p_r}
\end{equation}
maximizes $\mathcal{L}(D,\{p_r\})$. Therefore, according to the
\emph{maximum likelihood method} \cite{Casella2001}, with a fixed
$D$, it is easy to determine $\{p_r\}$ (by Eq. \ref{optimal p_r})
that best fits the network $G$. Figure \ref{hierarchical-inferring}
shows an example network and two possible dendrograms, as well as
the corresponding likelihoods. It is in accordance with the common
sense that $D_2$ is more likely. A \emph{Markov chain Monte Carlo
method} is used to sample dendrograms with probability proportional
to their likelihood (see the Supplementary Information of Ref.
\cite{Clauset2008} and a benchmark book \cite{Newman1999} for
details).

The algorithm to predict the missing links contains the following
procedures: (i) Sample a large number of dendrograms with
probability proportional to their likelihood; (ii) For each pair of
unconnected nodes $i$ and $j$, calculate the mean connecting
probability $\langle p_{ij}\rangle$ by averaging the corresponding
probability $p_{ij}$ over all sampled dendrograms; (iii) Sort these
node pairs in descending order of $\langle p_{ij}\rangle$ and the
highest-ranked ones are those to be predicted. According to the AUC
value, this algorithm outperforms the CN index for the terrorist
association network \cite{Krebs2002} and the grassland species food
web \cite{Dawah1995}, while loses for the metabolic network of the
spirochaete \emph{Treponema Pallidum} \cite{Huss2007}.

The hierarchical structure model provides a smart way to predict
missing links, and, maybe more significantly, it uncovers the hidden
hierarchical organization of networks. However, as mentioned above,
a big disadvantage is that this algorithm runs very slow. Actually,
the process to sample dendrograms usually asks for $\mathbb{O}(N^2)$
steps of the Markov chain \cite{Clauset2008}, and in the worst case,
it takes exponential time \cite{Mossel2005}. In comparison,
according to the CPU of an advanced desktop computer, the
hierarchical structure model cannot manage a network of tens of
thousands nodes, while the algorithms based on local similarity
indices can deal with networks with tens of millions nodes. Another
noticeable remark is that this model may give poor predictions for
those networks without clear hierarchical structures.

\subsection{Stochastic Block Model}

Stochastic block model
\cite{White1976,Holland1983,Dorelan2005,Airoldi2008} is one of the
most general network models, where nodes are partitioned into groups
and the probability that two nodes are connected depends solely on
the groups to which they belong. The stochastic block model can
capture the community structure \cite{Girvan2002}, role-to-role
connections \cite{Guimera2007} and maybe other factors for the
establishing of connections, especially when the group membership
plays the considerable roles in determining how nodes interact with
each other, which usually could not be well described by the simple
assortativity coefficient \cite{Newman2002,Newman2003b} or the
degree-degree correlations \cite{Pastor-Satorras2001,Vazquez2002}.

\begin{figure}
\centering
\includegraphics[scale=0.7]{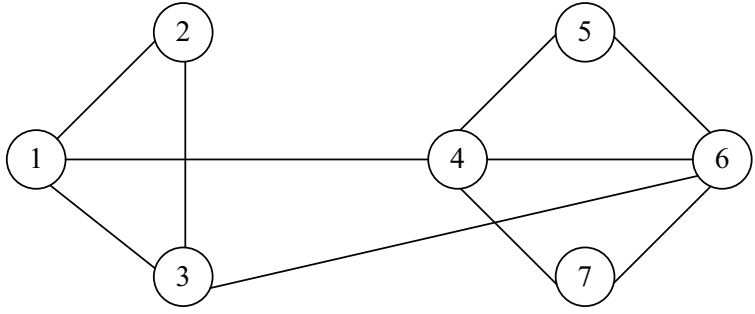}
\caption{An illustration about the calculation of likelihood for the
stochastic block model.} \label{stochastic block model}
\end{figure}

Given a partition $\mathcal{M}$ where each node belongs to one group
and the connecting probability for two nodes respectively in groups
$\alpha$ and $\beta$ is denoted by $Q_{\alpha \beta}$ ($Q_{\alpha
\alpha}$ represents the probability that two nodes within group
$\alpha$ are connected), then the likelihood of the observed network
structure is \cite{Guimera2009}:
\begin{equation}
\mathcal{L}(A|\mathcal{M})=\prod_{\alpha\leq \beta}Q_{\alpha
\beta}^{l_{\alpha \beta}}(1-Q_{\alpha \beta})^{r_{\alpha
\beta}-l_{\alpha \beta}}, \label{likelihood-SBM}
\end{equation}
where $l_{\alpha \beta}$ is the number of edges between nodes in
groups $\alpha$ and $\beta$ and $r_{\alpha \beta}$ is the number of
pairs of nodes such that one node is in $\alpha$ and the other is in
$\beta$. Similar to Eq. \ref{optimal p_r}, the optimal $Q_{\alpha
\beta}$ that maximizes the likelihood $\mathcal{L}(A|\mathcal{M})$
is:
\begin{equation}
Q_{\alpha \beta}^*=\frac{l_{\alpha \beta}}{r_{\alpha \beta}}.
\label{optimal Q}
\end{equation}

A simple illustration is shown in Fig. \ref{stochastic block model}.
Given a partition $\mathcal{M}=\{\{1,2,3\},\{4,5,6,7\}\}$, according
to Eq. \ref{optimal Q}, the $Q$ values corresponding to the maximum
likelihood are $Q_{11}^*=\frac{3}{3}=1$,
$Q_{12}^*=\frac{2}{12}=\frac{1}{6}$, $Q_{22}^*=\frac{5}{6}$, and
thus the likelihood is
\begin{equation}
\mathcal{L}=1\times
\left(\frac{1}{6}\right)^2\left(\frac{5}{6}\right)^{10}\times
\left(\frac{5}{6}\right)^5\left(\frac{1}{6}\right)\approx
3.005\times 10^{-4}. \label{likelihood calculation}
\end{equation}

Denote by $\Omega$ the set of all possible partitions, using
\emph{Bayes' Theorem} \cite{Bayes1763}, the \emph{reliability} of an
individual link is \cite{Guimera2009}:
\begin{equation}
R_{xy}=\mathcal{L}(A_{xy}=1|A)=\frac{\int_\Omega\mathcal{L}(A_{xy}=1|\mathcal{M})\mathcal{L}(A|\mathcal{M})p(\mathcal{M})\texttt{d}\mathcal{M}}{\int_\Omega\mathcal{L}(A|\mathcal{M}')p(\mathcal{M}')\texttt{d}\mathcal{M}'},
\label{reliability}
\end{equation}
where $p(\mathcal{M})$ is a constant assuming no prior knowledge
about the model. Notice that, the number of different partitions of
$N$ elements grows faster than any finite power of $N$, and thus
even for a small network, to sum over all partitions is not possible
in practice. The \emph{Metropolis algorithm} \cite{Metropolis1953}
can be applied to estimate the link reliability \cite{Guimera2009}.
Even though, the whole process is very time consuming and this
method can only manage networks with up to a few thousands of nodes.

Reliability describes the likelihood of the existence of a link
(i.e., the probability that the link ``truly" exists) given the
observed structure \cite{Guimera2009}, which can be used not only to
predict missing links (the nonexistent links in the observed network
yet with the highest reliabilities) but also to identify possible
spurious links (the existent links with the lowest reliabilities).
Empirical comparison on five disparate networks (the social
interactions in a karate club \cite{Zachary1977}, the social network
of frequent associations between 62 dolphins \cite{Lusseau2003}, the
air transportation network of Eastern Europe \cite{Guimera2005}, the
neural network of the nematode \emph{Caenorhabditis elegans}
\cite{White1986}, and the metabolic network of \emph{Escherichia
coli} \cite{Reed2003}) indicated that the overall performance of the
maximum likelihood method based on stochastic block model
\cite{Guimera2009} is better than the one based on the hierarchical
structure model \cite{Clauset2008} and the similarity-based
algorithm for common neighbors \cite{Liben-Nowell2007}.

\section{Probabilistic Models}
Probabilistic models aim at abstracting the underlying structure
from the observed network, and then predicting the missing links by
using the learned model. Given a target network $G=(V,E)$, the
probabilistic model will optimize a built target function to
establish a model composed of a group of parameters $\Theta$, which
can best fit the observed data of the target network. Then the
probability of the existence of a nonexistent link $(i,j)$ is
estimated by the conditional probability $P(A_{ij}=1|\Theta)$. This
section will introduce the three mainstream methods, respectively
called \emph{Probabilistic Relational Model} (PRM)
\cite{Friedman1999}, \emph{Probabilistic Entity Relationship Model}
(PERM) \cite{Heckerman2004_1} and Stochastic Relational Model (SRM)
\cite{Yu2006}. Notice that, in some literature, the term PRM only
refers to a specific model which is usually called the
\emph{Relational Bayesian Networks} nowadays, while we adopt the
more general usage of PRM in this review.

\subsection{Probabilistic Relational Models}
PRMs represent a joint probability distribution over the attributes
of a relational dataset. They allow the properties of an object to
dependent probabilistically both on other properties of that object
and on properties of the related objects. Different from the
traditional graphical models using a single graph to model the
relationship among the attributes of homogenous entities, PRMs
contain three graphs \cite{Neville2006}: the data graph $G_D$, the
model graph $G_M$, and the inference graph $G_I$. These correspond
to the skeleton, model, and ground graph as outlined by Heckerman
\emph{et al.} \cite{Heckerman2004_2}.

The data graph $G_D = (V_D,E_D)$ presents the input network, where
nodes are the objects in the data and edges represent the
relationships among the objects. Each node $v_i\in V_D$ and edge
$e_j\in E_D$ are associated with a type $T(v_i) = t_{v_i}$, $T(e_j)
= t_{e_j}$. Each item (either object or edge) type $t \in T$ has a
number of associated attributes $\mathrm{X}^t$. Consequently, each
object $v_i$ and link $e_j$ are associated with a set of attribute
values, $\mathrm{x}^{t_{v_i}}_{v_i}$ and
$\mathrm{x}^{t_{e_j}}_{e_j}$, determined by their types, $t_v$ and
$t_e$, respectively. A PRM represents a joint probability
distribution over the values of all the attributes in the data
graph, $\mathrm{x} = \{\mathrm{x}^{t_{v_i}}_{v_i} : v_i \in V_D,
T(v_i) = t_{v_i}\} \bigcup \{\mathrm{x}^{t_{e_j}}_{e_j} : e_j \in
E_D, T(e_j) = t_{e_j}\}$. For example, in the student-course
selection system, the students and courses are nodes and the edges
represent the \emph{select} relationship between students and
courses. Clearly, there are two types of nodes, namely student and
course. And the type \emph{student} has four attributes: grade, age,
sex and department, while the type \emph{course} has five
attributes: category, teacher, year, time and discipline.

The model graph $G_M=(V_M,E_M)$ represents the dependencies among
attributes at the level of item types. Attributes of an item can
depend probabilistically on other attributes of the same item, as
well as on attributes of other related objects or links in $G_D$.
Each node in $V_M$ corresponds to an attribute $X_i^t \in
\mathrm{X}^t$ where $t\in T$. The attributes with the same type in
$G_D$ are tied together. Thus $G_D$ is decomposed into multiple
examples of each type, based on which a joint model of dependencies
among the type attributes can be built. $G_M$ contains two parts:
the dependent structure among all the type attributes and the
conditional probability distribution (CPD) associated with the nodes
in $G_M$.

The inference graph $G_I=(V_I,E_I)$ represents the probabilistic
dependencies among all the variables in a single test set. It can be
instantiated by a roll-out process of $G_D$ and $G_M$. Each
item-attribute pair in $G_D$ gets a separate, local copy of the
corresponding CPD from $G_M$. The relations in $G_D$ determine the
way that $G_M$ is rolled out to form $G_I$. Therefore the structure
of $G_I$ is determined by both $G_D$ and $G_M$.

With respect to different representations of the modeled graph $G_M$
and the corresponding learning and inferring procedures, PRMs can be
classified into three groups: Relational Bayesian Networks (RBNs),
Relational Markov Networks (RMNs) and Relational Dependency Networks
(RDNs).

\emph{Relational Bayesian Networks} (RBNs)
\cite{Friedman1999,Heckerman1995}--- The model graph presented by a
RBN is a \emph{directed acyclic graph}\footnote{A directed graph is
acyclic if there is no directed path that starts and ends at the
same variable. This constrain indicates that a random variable does
not depend, directly or indirectly, on its own value.} with a set of
CPDs, $P$, to represent a joint distribution over the attributes of
the item types. The set $P$ contains a conditional probability
distribution for each variable given its parents\footnote{A direct
link from $a$ to $b$ indicates that $a$ is $b$'s parent node.},
$p(x|pa_x)$, where $pa_x$ denotes the parents of node. Thus the
joint probabilistic distribution can be calculated as
\begin{equation}
p(\mathrm{x})=\prod_{t\in T} \prod_{X_i^t\in
\mathrm{X}^t}\prod_{v:T(v)=t}
p(x^t_{v_i}|pa_{x^t_{v_i}})\prod_{e:T(e)=t}
p(x^t_{e_i}|pa_{x^t_{e_i}}).
\end{equation}
The need to avoid cycles in RBN leads to significant
representational and computational difficulties. Inference is done
by creating the complete ground network, which limits their
scalability. RBN requires specifying a complete conditional model
for each attribute of each class, which in large complex domains can
be quite burdensome.

\emph{Relational Markov Networks} (RMNs)
\cite{Taskar2002,Taskar2003}--- A RMN uses an undirected graph and a
set of potential functions $\Phi$ to represent the joint
distribution over the attributes of the item types. Denote by $C$
the set of cliques in the graph and each clique $c \in C$ is
associated with a set of variables $\mathrm{X}_c$ ($\in
\mathrm{X}^t$) (i.e. the nodes in this clique \footnote{Actually a
node in a clique corresponds to an attribute in the data graph.})
and a clique potential $\Phi_c(x_c)$ which is a non-negative
function over the possible values for $x_c\in\mathrm{X}_c$, then the
joint probability over $\mathrm{x}$ is calculated with the formula
\begin{equation}
p(\mathrm{x})=\frac{1}{Z}\prod_{c\in C} \Phi_c(x_c),
\end{equation}
where $Z$ is a normalizing constant, which sums over all possible
instantiations. RMNs are trained discriminatively, and do not
specify a complete joint distribution for the variables in the
model. The learning algorithm uses maximum a posteriori (MAP)
estimation with belief propagation for inference, which leads to a
high computational complexity for learning.

\emph{Relational Dependency Networks} (RDNs)
\cite{Heckerman2000,Neville2007}--- The RDN is a bi-directed
graphical model with a set of conditional probability distributions,
which can be used to represent the cyclic dependencies in a
relational setting. RDNs use pseudo-likelihood learning techniques
to estimate an efficient approximation of the full joint
distribution of the attribute values in a relational dataset. The
pseudo-likelihood for data graph $G_D$ is computed as a product over
the item types $t$, the attributes of that type $\mathrm{X}^t$, and
the nodes $v$ and edges $e$ of that type:
\begin{equation}
PL(G_D;\Theta)=\prod_{t\in T}\prod_{X_i^t\in
\mathrm{X}^t}\prod_{v:T(v)=t}
p(x_{v_i}^t|pa_{x_{v_i}^t};\Theta)\prod_{e:T(e)=t}p(x_{e_i}^t|pa_{x_{e_i}^t};\Theta).
\end{equation}
The CPDs in the RDN pseudo-likelihood are not required to factor the
joint distribution of $G_D$. More specifically, when consider the
variable $x_{v_i}^t$, we condition on the values of the parents
$pa_{x_{v_i}^t}$ regardless of whether the estimation of CPDs for
variables in $pa_{x_{v_i}^t}$ was conditioned on $x_{v_i}^t$. RDN
adopts \emph{Gibbs sampling}\footnote{Heckerman \emph{et al.}
\cite{Heckerman2000} proved that the Gibbs sampling can be used to
estimate the joint distribution of a dependency network. For a basic
introduction and summary of the Gibbs sampling, see Ref.
\cite{Casella1992}.} to iteratively relabel each unobserved variable
by drawing from its local conditional distribution, given the
current state of the rest of the graph.

\subsection{Probabilistic Entity Relationship Models}
A specific type of probabilistic entity-relationship model is the
directed acyclic PERM (DAPER for short), which uses directed
arcs\footnote{In graph theory, the term ``arc" stands for directed
link.} to describe the relationship between attributes
\cite{Heckerman2004_1}. DAPER makes relationships the first class
objects in the modeling language, and encourages an explicit
representation of conditional probabilistic distribution. The DAPER
model consists of six classes \cite{Xu2005}: (i) Entity classes:
specify the classes of objects in real world; (ii) Relationship
classes: represent the interaction among entity classes; (iii)
Attribute classes: describe properties of entities or relationships.
(iv) Arc classes: represent probabilistic dependencies among
corresponding attributes; (v) Local distribution classes: construct
the local distributions for attributes corresponding to the
attribute class; (vi) Constraint classes: specify how to derive
inference graph (i.e., ground graph) from the corresponding DAPER
model over the given instantiated domain. DAPER model assigns
relationships the same importance as the entities.

The DAPER model can be used in the situation where the relational
structure itself is uncertain. And it is more expressive than either
PRMs or \emph{plate models}\footnote{The standard description of
plate models can be found in Refs.
\cite{Buntine1994,Spiegelhalter1998}. Heckerman \emph{et al.}
\cite{Heckerman2004_1} provided a new definition of plate model,
which is slightly different from the traditional one
\cite{Buntine1994,Spiegelhalter1998}. According to this new
definition, the plate models and DAPER models are equivalent, and a
plate model can be invertible mapped to a DAPER model
\cite{Heckerman2004_1}.}. Actually, DAPER combines the features of
plate models and PRMs, and the relations between DAPER models, PRMs
and plate models can be found in Ref. \cite{Heckerman2004_1}.

\subsection{Stochastic Relational Models}
The key idea of SRM is to model the stochastic structure of entity
relationships (i.e., links) via a tensor interaction of multiple
Gaussian Processes (GPs), each defined on one type of entities
\cite{Yu2006}.

Assuming that the observable links $r$ are derived as local
measurements of a real-value latent relation function $t$:
$\mathcal{U}\times \mathcal{V}\rightarrow \mathbb{R}$, and each link
$r_{ij}$ is solely dependent on its latent value $t_{ij}$, modeled
by the likelihood $p(r_{ij}|t_{ij})$. The latent relation function
$t$ is generated via a tensor interaction of two independent
entity-specific GPs, one acting on $\mathcal{U}$ and the other on
$\mathcal{V}$. Note that $\mathcal{U}$ and $\mathcal{V}$ can both
encompass an infinite number of entities. The relational processes
are characterized by hyper-parameters
$\Theta=\{\Theta_\Sigma,\Theta_\Omega\}$, where $\Theta_\Sigma$ and
$\Theta_\Omega$ are for the GP kernel function on $\mathcal{U}$ and
$\mathcal{V}$ respectively. Thus SRM defines a Bayesian prior
$p(t|\Theta)$ for the latent variables $t$. Let $\mathbb{I}$ be the
index set of entity pairs having observed links, the marginal
likelihood under such a prior is
\begin{equation}
p(\mathbf{R}_{\mathbb{I}}|\Theta)=\int\prod_{(ij)\in \mathbb{I}}
p(r_{ij}|t_{ij})p(t|\Theta)dt,
\end{equation}
where $\mathbf{R}_{\mathbb{I}}={r_{ij}}$, ${(i,j)\in\mathbb{I}}$.
The hyper-parameters $\Theta$ can be estimated by maximizing the
marginal likelihood. Then the link for a new pair of entities can be
predicted by marginalizing over the \emph{a posteriori}
$p(t|\mathbf{R}_{\mathbb{I}},\Theta)$.

This model in fact defines a set of nonparametric priors on infinite
dimensional tensor matrices, where each element represents a
relationship between a tuple of entities. By maximizing the
marginalized likelihood, information is exchanged between the
participating GPs through the entire relational network, so that the
dependency structure of links is messaged to the dependency of
entities, reflected by the adapted GP kernels. Because the training
is on a conditional model of links, this model offers a
discriminative approach for link prediction, namely predicting the
existences, strengths, or types of relationships based on the
partially observed linkage network as well as the attributes of
entities if given. Yu \emph{et al.} further upgraded SRM with an
edge-wise covariance with which the overall computational complexity
can be reduced \cite{Yu2007}. For more details one can see Refs.
\cite{Yu2006,Chu2006}.

\section{Applications}

The problem of link prediction has attracted much attention from
disparate research communities. This is mainly attributed to its
broad applicability. For some networks, especially biological
networks such as protein-protein interaction networks, metabolic
networks and food webs, the discovery of links or interactions is
costly in the laboratory or the field. A highly accurate prediction
can reduce the experimental costs and speed the pace of uncovering
the truth \cite{Clauset2008,Redner2008}. Link prediction has also
been applied in the analysis of social networks, such as the
prediction of being actors in acts \cite{Madadhain2005}, the
prediction of the collaborations in co-authorship networks
\cite{Liben-Nowell2007}, the detection of the underground
relationships between terrorists \cite{Clauset2008}, and so on. In
addition, the process of recommending items to users can be
considered as a link prediction problem in the user-item bipartite
networks \cite{Shang2010,Kunegis2010}. Actually, almost the same
techniques as the similarity-based link prediction has been applied
in personalized recommendation \cite{Zhou2007,Zhou2010,Zeng2010}.
Accurate recommendation can be used in e-commerce web sites to
enhance the sales \cite{Schafer2001}. Moreover, the link prediction
approaches can be applied to solve the classification problem in
partially labeled networks, such as the prediction of protein
functions \cite{Holme2005}, the detection of anomalous email
\cite{Huang2006}, distinguishing the research ares of scientific
publications \cite{Gallagher2008}, and finding out the fraud and
legit users in cell phone networks \cite{Dasgupta2008}. The
following three subsections will introduce typical applications of
link prediction.

\subsection{Reconstruction of Networks}
Guimer\`a and Sales-Pardo \cite{Guimera2009} firstly considered the
reconstruction of networks from the observed networks with missing
and spurious links. Although one can rank the observed and
non-observed links according to their reliabilities (see Eq.
\ref{reliability}), it is not easy to reconstruct the ``true"
network since generally no one knows how many missing and spurious
links there are. Applying the similar techniques presented in
Section 4.2, Guimer\`a and Sales-Pardo \cite{Guimera2009} defined
the reliability of a network $A$ as
\begin{equation}
R(A)=\prod_{A_{xy}=1,x<y}R_{xy}=\prod_{A_{xy}=1,x<y}\mathcal{L}(A_{xy}=1|A^O),
\label{Network Reliability}
\end{equation}
where $R_{xy}$ and $\mathcal{L}$ are defined in Eqs.
\ref{likelihood-SBM} and \ref{reliability}, and the term $A^O$ is
used to emphasize that the likelihoods are calculated according to
the observed network.

Given $A^O$, a straightforward idea is to find out the network $A$
that maximizes the reliability defined by Eq. \ref{Network
Reliability}. However, the computation is too costly to be
implemented. In practice, Guimer\`a and Sales-Pardo
\cite{Guimera2009} designed a simple greedy algorithm. Their
algorithm starts by evaluating the link reliabilities for all
pairs of nodes. Then, at each time step it removes the link with
the lowest reliability and adds the link (not yet in the current
network) with the highest reliability. This change is accepted if
and only if the network reliability increases. If it is rejected,
the link with the next lowest reliability and the not-yet-existent
link with the next highest reliability will be the next candidate
for swapping. The algorithm stops if it rejects five consecutive
attempts to swap links. The observed network is set as the
initialization of the algorithm, and it will consecutively become
another network with higher reliability than the initial network.
Guimer\`a and Sales-Pardo \cite{Guimera2009} tested their
algorithm by generating hypothetical observed networks $A^O$ from
the true networks $A^T$ (the five true networks used for testing
are introduced in Section 4.2). Each observation has a fraction of
the true links removed and an identical number of random links
added.

Guimer\`a and Sales-Pardo \cite{Guimera2009} compared the global
network properties of the observed networks and these of the
reconstructed networks. According to six metrics, clustering
coefficient \cite{Watts1998}, modularity \cite{Newman2004},
assortativity \cite{Newman2002,Newman2003b},
congestability\footnote{The congestability refers to the maximal
betweenness centrality which governs the transportation throughput
of a network \cite{Guimera2002,Yan2006}.},
synchronizability\footnote{The synchronizability refers to the
ratio between the largest and the smallest non-zero eigenvalues of
the Laplacian matrix of a network, which quantifies the ability of
synchronization under the framework of master stability analysis
\cite{Barahona2002,Arenas2008}.} and spreading
threshold\footnote{Ignoring the degree-degree correlations and
applying the mean-field approximation, the spreading threshold
equals the ratio between the first and the second moments of the
degree distribution \cite{Pastor-Satorras2003,Zhou2006b}.}, the
reconstruction generally improves the estimates, indicating the
validity of the approach. Notice that, the results from the greedy
algorithm may be far different from the real optimum subject to
the maximal reliability, thus we may expect even better estimates
if one has developed a more effective and/or efficient algorithm.
Readers should be warned that in both the algorithm and the
preparation of observed networks, a latent assumption is that the
number of missing links and the number of spurious links are
equal. Since in the real systems, these two numbers may be very
different (it is easy to image that in many networks, such as
metabolic networks and friendship networks, the missing links are
much more than the spurious links), the effectiveness of the
algorithm still needs further validation.

\subsection{Evaluation of Network Evolving Mechanisms}

Since the groundbreaking work by Barab\'asi and Albert
\cite{Barabasi1999}, the evolving models all the time lie in the
center of the complex network study. A fundamental difficulty is
that for a given network or a target network property, there are
generally many possible mechanisms and it is not easy to judge which
one is the best. Taking the power-law degree distribution as an
instance, the well-known mechanisms include \emph{rich gets richer}
\cite{Barabasi1999}, \emph{good gets richer} \cite{Caldarelli2002},
\emph{optimal design} \cite{Valverde2002}, \emph{Hamiltonian
dynamics} \cite{Baiesi2003}, \emph{merging and regeneration}
\cite{Kim2005}, \emph{stability constraints} \cite{Perotti2009}, and
so on. Hence we can not easily know which factor(s) leads to the
scale-free property of a real network, and in fact there can be so
many models competing for the final explanation of a given real
network. It is very hard to evaluate different models by comparing
their resulted networks with the target network, since there are too
many metrics for topological features \cite{Costa2007}. As mentioned
in Section 1, there are many models about the topology of the
Internet, some more accurately reproduce the degree distribution and
the disassortative mixing pattern (e.g., see Ref. \cite{Zhou2004})
and some better characterize the $k$-core structure (e.g., see Ref.
\cite{Carmi2007}). To judge which model (i.e., which evolving
mechanism) is better than the others is a tough task.

Essentially speaking, an algorithm for link prediction makes a guess
about the factors resulting in the existence of links, which is
actually what an evolving model wants to show. In other words, an
evolving model in principle can be mapped to a link prediction
algorithm. Therefore, we can quantitatively compare the accuracies
of different evolving models with the help of the performance
metrics for link prediction (see Section 2). We hope this
methodology could provide a fair platform to compare different
evolving models, which may be significant for the studies of network
modeling. Next, we will show a real application about the Chinese
city airline network, where each node represents a city with
airport, and two cities are connected if there exists at least one
direct airline between them \cite{Liu2007}.

It is well-known that the evolution of a city airline network is
affected by not only the topological factor, but also the
geographical factor \cite{Bianconi2009} and external factors, such
as population and economic level of a city \cite{Liu2008}. As
shown by Liben-Nowell \emph{et al.} \cite{Liben-Nowell2007} and
Zhou \emph{et al.} \cite{Zhou2009}, the common neighbor index is a
good candidate to account for the topological effects. In
addition, Cui \emph{et al.} \cite{Cui2010} developed an evolving
model driven completely by the common neighborhood, which well
reproduces not only the global network properties, but also the
local structural features like power-law clique-degree
distributions \cite{Xiao2007} of social and technological
networks. Therefore, we simply use the common neighbor index
$S^{CN}$ (see Eq. \ref{Sim1}) to represent the topological
ingredient. Geographical distance is considered to be one of the
realistic factors that affect the existence of nodes' interactions
in networks \cite{Lambiotte2008}. Especially it plays very
important role in analyzing transportation networks
\cite{Jung2008,Kaluza2010}. It is known to be relevant to the
existence of an airline, and the number of airlines decays with
the increasing of corresponding distance
\cite{Gastner2006,Bianconi2009}. Accordingly, we use the inverse
of geographical distance between two cities as the similarity
index, say
\begin{equation}
s_{xy}^{DIS}=\frac{1}{D_g(x,y)}, \label{Sim21}
\end{equation}
where $D_g(x,y)$ denotes the geographical distance between cities
$x$ and $y$. Based on a null assumption that people in every city
have the same frequency of air travels, the similar index for
populations is defined as
\begin{equation}
s_{xy}^{POPU}=P(x)\times P(y), \label{Sim22}
\end{equation}
where $P(x)$ is the population of city $x$. The economic level of a
city can be roughly quantified by its \emph{gross domestic product}
(GDP)\footnote{The GDP is a measure of a city's overall economic
output. It is the market value of all final goods and services made
within the borders of a city in a year. Here we use the data of the
year 2005.}, and thus the corresponding similarity is defined as
\begin{equation}
s_{xy}^{GDP}=G(x)\times G(y), \label{Sim23}
\end{equation}
where $G(x)$ denotes the GDP of city $x$. Considering that the
airline business is most tightly related to the service industry,
besides the simple GDP, we user the third sector of GDP, named the
\emph{tertiary industry}\footnote{The tertiary industry (also called
\emph{tertiary sector of the economy}, \emph{service sector} or
\emph{service industry}) consists of the ``soft" parts of the
economy, namely activities where people offer their knowledge and
time to improve productivity, performance, potential, and
sustainability. The basic characteristic of this sector is the
production of services instead of end products.} to characterize a
city's potential to build airlines:
\begin{equation}
s_{xy}^{TI}=T(x)\times T(y), \label{Sim24}
\end{equation}
where $T(x)$ is the tertiary industry of city $x$.

\begin{table}
\caption{The prediction accuracy of the five similarity indices for
the Chinese city airline network. The training and testing sets are
divided according to the leave-one-out method.}
\begin{center}
\begin{tabular} {cc}
  \hline \hline
   Similarity Indices  & AUC \\
   \hline
$S^{CN}$     & 0.898 \\
$S^{DIS}$     & 0.699 \\
$S^{POPU}$     & 0.745 \\
$S^{GDP}$     & 0.855 \\
$S^{TI}$     & 0.881 \\
   \hline \hline
\end{tabular}
\end{center}
\label{Airline Network}
\end{table}

Since the size of the Chinese city airline network \cite{Liu2007}
is small ($|V|=121$, $|E|=1378$), we adopt the
\emph{leave-one-out} method, namely at each time, we pick only one
link for test and all other links constitute the training set.
This procedure repeats for 1378 times with each link being once
the testing link. Table \ref{Airline Network} displays the
prediction accuracy (AUC values) of the five similarity indices.
It indicates that every factor under consideration plays a role,
while the topological factor is most significant. The tertiary
industry of a city, as an external factor, also plays a very
important role. Actually, a linear combination of the common
neighbor index and the tertiary industry, as $S'=\lambda
S^{CN}+(1-\lambda)S^{TI}$ can achieve a very high AUC value,
0.928, at $\lambda\approx 0.2$.

Although the method introduced here is straightforward, it gives
insights into the underlying evolving mechanisms which may not be
seriously considered in the early studies. The validity of this
method has been demonstrated by some recent evidences. For example,
by comparing the evolving models driven respectively by the
topological factor, the geographical factor, and the above-mentioned
three external factors, Liu \emph{et al.} \cite{Liu2009} showed that
only the one considering the tertiary industry can reproduce the
observed double power-law degree distribution of the Chinese city
airline network. In addition, among many external factors, the
\emph{Granger causality test}\footnote{The Granger causality test is
a technique for determining whether one time series is useful in
forecasting another. See Ref. \cite{Granger1969} for details.} shows
that the tertiary industry is the most significant factor in
determining the passenger volume \cite{Liu2010b}.

\subsection{Classification of Partially Labeled Networks}

Given a network with partial nodes being labeled, the problem is to
predict the labels of these unlabeled nodes based on the known
labels and the network structure. Two main difficulties in achieving
highly accurate classification are the sparsity of the known labeled
nodes and the inconsistency of label information. To address these
two difficulties, a simple but effective method is to add artificial
connections between every pair of labeled and unlabeled nodes
according to their similarity scores
\cite{Gallagher2008,ZhangQM2010}, with almost the same techniques
used in similarity-based link prediction. An underlying assumption
is that two nodes are more likely to be categorized into the same
class if they are more similar to each other.

\begin{figure}
\begin{center}
\includegraphics[width=5cm]{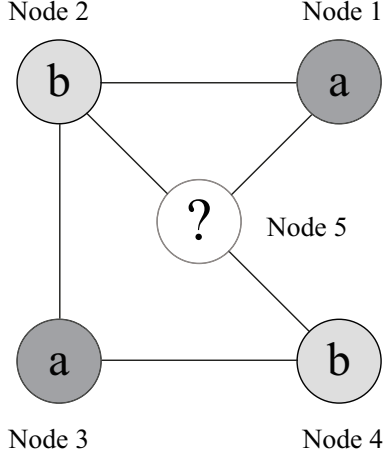}
\caption{An illustration of how to predict the fifth node's label by
adding artificial links.\label{Class}}
\end{center}
\end{figure}

Consider an unweighted undirected network of both labeled and
unlabeled nodes: $G(V, E, L)$, where $V$ is the set of nodes, $E$ is
the set of links and $L=\{l_1,l_2,\cdots,l_m\}$ is the set of
labels. For the nodes without labels, we label them by 0. For each
pair of nodes, $x$ and $y$, a similarity index will assign a score
as $s_{xy}$. For an unlabeled node $x$, the probability it belongs
to $l_i$ is
\begin{equation}
p(l_i|x)=\frac{\sum_{\{y|y\neq{x},label(y)=l_i\}}s_{xy}}{\sum_{\{y|y\neq{x},label(y)\neq
0\}}s_{xy}},
\end{equation}
where $l_i\in{L}$. The predicted label of node $x$ is determined by
the largest $p(l_i|x)$. If there are more than one maximum values,
we randomly select one.

A simple example is shown in Fig.~\ref{Class}, where there are two
kinds of labels (i.e. $a$ and $b$) and five nodes, four of which
are labeled already. Our task is to predict the label of the node
5. According to the common neighbors index $S^{CN}$, we obtain the
similarity between node 5 and the other four labeled nodes:
$s_{15}=1$, $s_{25}=1$, $s_{35}=2$ and $s_{45}=0$. Thus, the
probabilities that node 5 belongs to class $a$ and $b$ are
$p(a|\texttt{node}5)=0.75$ and $p(b|\texttt{node}5)=0.25$
respectively. If we use RA index, the similarity scores are:
$s_{15}=\frac{1}{3}$, $s_{25}=\frac{1}{2}$,
$s_{35}=\frac{1}{3}+\frac{1}{2}$ and $s_{45}=0$. Therefore, the
probabilities change to $p(a|\texttt{node}5)=0.7$ and
$p(b|\texttt{node}5)=0.3$. According to any of the two indices,
the predicted label of node 5 is $a$.

\section{Outlook}

In this article, we briefly summarized the progress of studies on
link prediction, emphasizing on the recent contributions by
statistical physicists. Although link prediction is not a new
problem in information science, traditional methods have not caught
up the new development of network science, especially the new
perspectives and tools resulted from the studies of complex
networks. In our opinion, the studies of link prediction and complex
networks will benefit each other, because the in-depth understanding
of network structure can be used to design advanced link prediction
algorithms (e.g., making use of the information about hierarchical
organization \cite{Clauset2008} and modular structure
\cite{Guimera2009} of real networks to better predict missing links)
and the performance of a link prediction algorithm could give
evidences about structural features \cite{Zhou2009} as well as the
algorithms themselves can be used to improve the estimates of real
networks' properties \cite{Guimera2009} and to evaluate the evolving
network models. In a word, to statistical physicists, the study of
link prediction is just unfolding.

Up to now, the studies of link prediction overwhelmingly emphasize
on the unweighted undirected networks. For directed networks, even
the ternary relations are complicated, and thus the simple
common-neighbor-based similarity indices have to be modified to
take into account the local motif structure \cite{Alon2007}.
Otherwise even we can predict the existence of an arc between two
nodes, we can not determine its direction. In addition, the
path-dependent similarity indices should also be extended to take
into account the link direction \cite{Mantrach2010b}. The
fundamental task of link prediction in weighted networks, namely
to predict the existence of links with the help of not only the
observed links but also their weights, has already been considered
by Murata \emph{et al.} \cite{Murata2007} and L\"u \emph{et al.}
\cite{Lu2010}. The former \cite{Murata2007} suggested that the
links with higher weights are more important in predicting missing
links, while the latter \cite{Lu2010} indicated a completely
opposite conclusion: the weak links play a more significant role.
How to properly exploit the information of weights to improve the
prediction accuracy is still an unsolved problem. A harder problem
is to predict the weights of links, which is relevant to the
traffic prediction for urban transportation and air transportation
systems \cite{Yin2002}. We are expecting some variants of link
prediction algorithms can also contribute to this domain.

A big challenge is the link prediction in multi-dimensional
networks, where links could have different meanings. For example, a
social network may consist of positive and negative links,
respectively pointing to friends and foes \cite{Kunegis2009}, or
trusted and distrusted peers \cite{Guha2004}. Leskovec \emph{et al.}
\cite{Leskovec2010} proposed a method to predict the signs of links
(positive or negative), yet the prediction of both the existence of
a link and its sign has not been well studied. Recent development of
\emph{social balance theory} may provide useful hints
\cite{Traag2009,Marvel2009,Szell2010}.

A more complicated kind of multi-dimensional networks is the ones
consisted of several classes of nodes. For example, an online
resource-sharing system, such as Del.icio.us\footnote{Del.icio.us is
the largest social bookmarking system where a user is allowed to
collect URLs as well as visit and impose other users' collections.},
can be represented by a network that consists of three kinds of
nodes: users, URLs and tags. Different from the tripartite networks,
nodes in the same class can also be connected, like an arc can be
added from a user to her/his follower who has imposed her/his
collections. Ignoring the connections within a class of nodes, the
prediction of links between users and objects has already been
investigated \cite{Zhang2010}. However, there is still nothing
reported about the link prediction algorithms taking into account
both the links within a class and the links between classes.

Inspired by the success in recommender systems, we think the
prediction accuracy can be considerably improved by hybrid
algorithms \cite{Burke2002}. Given a specific target network, we can
implement many individual prediction algorithms, and then try to
select and organize them in a property way. This so-called
\emph{ensemble learning} method can obtain better prediction
performance than could be obtained from any of the individual
algorithms \cite{Polikar2006}. Although the scientific significance
of such a method is not clear to us, building ensemble systems for
link prediction could be of huge practical value.

The algorithms' performance can be effectively enhanced by
considering some external information, like the attributes of
nodes \cite{Lin1998}. In common sense, two people share more
tastes and interests (and thus may of higher probability to be
connected in a social network) if they have more common features,
such as age, sex, job, and so on. The attribute information can be
used to predict links without considering the network structures.
Thus, when the existed links themselves are unreliable,
attribute-based methods are preferable, which can to some extent
solve the so-called cold start problem -- a big challenge of link
prediction \cite{Leroy2010}. Besides, community structures can
also help improving prediction accuracy \cite{Zheleva2008}. In
social networks, since one person may play different roles in
different communities, the prediction in one domain can be
inspired by the information in others \cite{Cao2010}. For example,
when we predict the collaborations between authors, we can
consider their affiliations to improve the accuracy.

Most of current approaches take into account a single snapshot of
a network to predict the missing or future links. Extensive
experiments show that these methods well uncover whether a link
exists. However, this static graph representation is difficult in
predicting the repeated link occurrences. For example, it is
impossible to predict whether and when two authors will
collaborate again in co-authorship network. Addressing this
problem, Huang and Lin \cite{Huang2009} proposed a time-series
link prediction approach considering the temporal evolutions of
link occurrences, which is more appropriate for dealing with the
link prediction problem in evolving networks, such as online
social networks. Another way to involve time information is
inspired by the fact that older events are less likely to be
relevant to future links than recent ones. For example, author's
interests may change over time and thus old publications might be
less relevant to his currents research area. Tylenda \emph{et al.}
\cite{Tylenda2009} developed a graph-based link prediction method
that incorporate the temporal information contained in evolving
networks. They found that the performance can be improved by
either time-based weighting of edges (i.e., giving the older
events smaller weights or even neglecting them) or weighting of
edges according to the connecting strength. However, to design
effective algorithms and eventually settle this problem, we need
in-depth and comprehensive understanding of temporal effects on
human's interests, attentions and so on, which asks for extensive
empirical analyses.

\section*{Acknowledgements}

We acknowledge Ci-Hang Jin, Hong-Kun Liu, Wei-Ping Liu, Ming-Sheng
Shang, Qian-Ming Zhang and Yi-Cheng Zhang for their contributions
to the collaborated works on link prediction, as well as Renaud
Lambiotte, Medo Matu\v{s}, Roger Guimer\`a, Marco Saerens, Marta
Sales-Pardo, Chi Ho Yeung and Zi-Ke Zhang for their valuable
discussions, comments and suggestions. This work is partially
supported by the National Natural Science Foundation of China
under Grant Nos. 11075031 and 10635040, and the Swiss National
Science Foundation under Grant No. 200020-121848.


\begin{thebibliography}{0}

\bibitem{Albert2002} R. Albert, A.-L. Barab\'asi, Statistical mechanics of complex networks, Rev. Mod. Phys.
74 (2002) 47.
\bibitem{Dorogovtsev2002} S. N. Dorogovtsev, J. F. F. Mendes, Evolution of networks, Adv.
Phys. 51 (2002) 1079.
\bibitem{Newman2003} M. E. J. Newman, The Structure and Function of
Complex Networks, SIAM Rev. 45 (2003) 167.
\bibitem{Boccaletti2006} S. Boccaletti, V. Latora, Y. Moreno, M.
Chavez, D.-U. Huang, Complex networks: Structure and dynamics, Phys.
Rep. 424 (2006) 175.
\bibitem{Costa2007} L. da F. Costa, F. A. Rodrigues, G. Travieso, P.
R. U. Boas, Characterization of complex networks: A survey of
measurements, Adv. Phys. 56 (2007) 167.
\bibitem{Salton1983} G. Salton, M. J. McGill, Introduction to Modern
Information Retrieval, MuGraw-Hill, Auckland, 1983.
\bibitem{Salton1989} G. Salton, Automatic Text Processing: The Transformation, Analysis,
and Retrieval of Information by Computer, Addison-Wesley, Boston,
1989.
\bibitem{Manning2008} C. D. Manning, P. Raghavan, H. Sch\"utze, Introduction to
Information Retrieval, Cambridge University Press, New York, 2008.
\bibitem{Getoor2005} L. Getoor, C. P. Diehl, Link Mining: A Survey, ACM SIGKDD
Explorations Newsletter 7 (2005) 3.
\bibitem{Yu2008} H. Yu \emph{et al.}, High-quality binary protein interaction map of
the yeast interactome network, Science 322 (2008) 104.
\bibitem{Stumpf2008} M. P. H. Stumpf, T. Thorne, E. de Silva, R. Stewart, H. J. An, M.
Lappe, C. Wiuf, Estimating the size of the human interactome, Proc.
Natl. Acad. Sci. U.S.A. 105 (2008) 6959.
\bibitem{Amaral2008} L. A. N. Amaral, A truer measure of our ignorance, Proc. Natl. Acad. Sci. U.S.A. 105 (2008) 6795.
\bibitem{Schafer2002} L. Schafer, J. W. Graham, Missing data: Our view of the state of the
art, Psychol. Methods 7 (2002) 147.
\bibitem{Kossinets2006} G. Kossinets, Effects of missing data in social networks,
Social Networks 28 (2006) 247.
\bibitem{Neal2008} J. W. Neal, ``Kracking" the Missing Data Problem: Applying Krackhardt's Cognitive Social Structures to School-Based Social Networks, Sociol. Educ. 81 (2008) 140.
\bibitem{Mering2002} C. von Mering, R. Krause, B. Snel, M. Cornell, S. G. Oliver, S.
Field, P. Bork, Comparative assessment of large-scale data sets of
protein-protein interactions, Nature 417 (2002) 399.
\bibitem{Butts2003} C. T. Butts, Network inference, error, and
information (in)accuracy: A Bayesian approach, Social Networks 25
(2003) 103.
\bibitem{Guimera2009} R. Guimer\`a, M. Sales-Pardo, Missing and spurious interactions and
the reconstruction of complex networks, Proc. Natl. Acad. Sci.
U.S.A. 106 (2009) 22073.
\bibitem{Zhou2004} S. Zhou, R. J. Mondrag\'on, Accurately modeling the internet
topology, Phys. Rev. E 70 (2004) 066108.
\bibitem{Carmi2007} S. Carmi, S. Havlin, S. Kirkpatrick, Y. Shavitt,
E. Shir, A model of Internet topology using k-shell decomposition,
Proc. Natl. Acad. Sci. U.S.A. 104 (2007) 11150.
\bibitem{Sales-Pardo2007} M. Sales-Pardo, R. Guimer\`a, L. A. N.
Amaral, Extracting the hierarchical organization of complex systems,
Proc. Natl. Acad. Sci. U.S.A. 104 (2007) 15224.
\bibitem{Girvan2002} M. Girvan, M. E. J. Newman, Community structure
in social and biological networks, Proc. Natl. Acad. Sci. U.S.A. 99
(2002) 7821.
\bibitem{Taylor2004} J. Shawe-Taylor, N. Cristianini, Kernels
Methods for Pattern Analysis, Cambridge University Press, Cambridge,
UK, 2004.
\bibitem{Breiman1992} L. Breiman, P. Spector, Submodel selection and
evaluation in regression: the $x$-random case, Int. Stat. Rev. 60
(1992) 291.
\bibitem{Kohavi1995} R. Kohavi, A Study of Cross-Valisation and
Bootstrap for Accuracy Estimation and Model Selection, In
Proceedings of the International Joint Conference on Artificial
Intelligence, Morgan Kaufmann Publisher, Quebec, Canada, 1995, p.
1137-1143.
\bibitem{Hanely1982} J. A. Hanely, B. J. McNeil, The meaning and use of the area under a receiver operating characteristic (ROC) curve, Radiology 143 (1982) 29.
\bibitem{Geisser1993} S. Geisser, Predictive inference: An introduction, Chapman
and Hall, New York, 1993.
\bibitem{Herlocker2004} J. L. Herlocker, J. A. Konstann, K. Terveen, J. T. Riedl, Evaluating collaborative filtering recommender systems, ACM Trans. Inf. Syst. 22 (2004) 5.
\bibitem{Su2009} X. Su, T. M. Khoshgoftaar, A Survey of
Collaborative Filtering Techniques, Advances in Artificial
Intelligence (2009) 421425.
\bibitem{Huang2005} Z. Huang, X. Li, H. Chen, Link prediction approach to collaborative filtering,
In Proceedings of the 5th ACM/IEEE-CS joint conference on Digital
libraries, ACM Press, New York, 2005.
\bibitem{Lin1998} D. Lin, An information-theoretic definition of similarity, in Proceedings of the
15th International Conference on Machine Learning, Morgan Kaufman
Publishers, San Francisco, 1998.
\bibitem{Leicht2006} E. A. Leicht, P. Holme, M. E. J. Newman, Vertex similarity in networks, Phys. Rev. E 73 (2006) 026120.
\bibitem{Sun2009} D. Sun, T. Zhou, J.-G. Liu, R.-R. Liu, C.-X. Jia,
B.-H. Wang, Information filtering based on transferring similarity,
Phys. Rev. E 80 (2009) 017101.
\bibitem{White1983} D. R. White, K. P. Reitz, Graph and semigroup homomorphisms on networks of relations, Social Networks 5 (1983) 193.
\bibitem{Holme2005} P. Holme, M. Huss, Role-similarity based functional prediction in networked systems: application to the yeast proteome, J. R. Soc. Interface 2
(2005) 327.
\bibitem{Newman2001a} M. E. J. Newman, Clustering and preferential attachment in growing networks, Phys. Rev. E 64 (2001) 025102.
\bibitem{Jaccard1901} P. Jaccard, \'{E}tude comparative de la distribution florale dans une portion des Alpes et des Jura, Bulletin de la Societe Vaudoise des Science Naturelles 37 (1901)
547.
\bibitem{Sorensen1948} T. S{\o}rensen, A method of establishing groups of equal amplitude in plant sociology based on similarity of species
content and its application to analyses of the vegetation on Danish
commons, Biol. Skr. 5 (1948) 1.
\bibitem{Ravasz2002} E. Ravasz, A. L. Somera, D. A. Mongru, Z. N.
Oltvai, A.-L. Barab\'asi, Hierarchical organization of modularity in
metabolic networks, Science 297 (2002) 1551.
\bibitem{Molloy1995} M. Molloy, B. Reed, A critical point for random graphs with a given degree sequence, Random Structure and Algorithms 6 (1995) 161.
\bibitem{Barabasi1999} A.-L. Barab\'asi, R.
Albert, Emergence of Scaling in Random Networks, Science 286 (1999)
509.
\bibitem{Xie2008} Y.-B. Xie, T. Zhou, B.-H. Wang, Scale-free networks without growth, Physica A 387 (2008) 1683.
\bibitem{Holme2002} P. Holme, B. J. Kim, C. N. Yoon, S. K. Han,
Attack vulnerability of complex networks, Phys. Rev. E 65 (2002)
056109.
\bibitem{Yin2006} C.-Y. Yin, W.-X. Wang, G.-R. Chen, B.-H. Wang, Decoupling process for better synchronizability on scale-free networks, Phys.
Rev. E 74 (2006) 047102.
\bibitem{Zhang2007} G.-Q. Zhang, D. Wang, G.-J. Li, Enhancing the transmission efficiency by edge deletion in scale-free networks, Phys. Rev. E 76 (2007) 017101.
\bibitem{Adamic2003} L. A. Adamic, E. Adar, Friends and neighbors on the Web, Social Networks 25 (2003) 211.
\bibitem{Zhou2009} T. Zhou, L. L\"u, Y.-C. Zhang, Predicting missing
links via local information, Eur. Phys. J. B 71 (2009) 623.
\bibitem{Ou2007} Q. Ou, Y.-D. Jin, T. Zhou, B.-H. Wang, B.-Q. Yin, Power-law strength-degree correlation from resource-allocation dynamics on weighted
networks, Phys. Rev. E 75 (2007) 021102.
\bibitem{Liben-Nowell2007} D. Liben-Nowell, J. Kleinberg, The link-prediction problem for social networks, J. Am. Soc.
Inform. Sci. Technol. 58 (2007) 1019.
\bibitem{Newman2006} M. E. J. Newman, Finding community structure in networks using the eigenvectors of matrices, Phys. Rev. E 74 (2006) 036104.
\bibitem{Watts1998} D. J. Watts and S. H. Strogatz, Collective dynamics of 'small-world' networks, Nature 393 (1998) 440 .
\bibitem{Ackland2005} R. Ackland, \emph{Mapping the US political
blogosphere: Are conservative bloggers more prominent},
\emph{Presentation to BlogTalk Downunder}, Sydney, 2005, available
at http://incsub.org/blogtalk/images/robertackland.pdf.
\bibitem{Spring2004} N. Spring, R. Mahajan, D. Wetherall and T.
Anderson, IEEE/ACM Trans. Networking 12 (2004) 2.
\bibitem{Batageli} V. Batageli, A. Mrvar, \emph{Pajek Datasets}, available
at http://vlado.fmf.uni-lj.si/pub/networks/data/default.htm.
\bibitem{Gastner2006} M. T. Gastner, M. E. J. Newman, The spatial structure of networks, Eur. Phys. J.
B 49 (2006) 247.
\bibitem{Liu2007} H.-K. Liu, T. Zhou, Empirical study of Chinese city airline network, Acta Physica Sinica 56 (2007)
106.
\bibitem{Zhou2004b} S. Zhou, R. J. Mondrag\'on, The rich-club phenomenon in the Internet topology, IEEE Commun. Lett. 8
(2004) 180.
\bibitem{Colizza2006} V. Colizza, A. Flammini, M. A. Serrano, A.
Vespignani, Detecting rich-club ordering in complex networks, Nat.
Phys. 2 (2006) 110.
\bibitem{Zhu2010} Y.-X. Zhu, L. L\"u, T. Zhou, Uncovering missing
links with cold ends (unpublished).
\bibitem{Pan2010} Y. Pan, D.-H. Li, J.-G. Liu, J.-Z. Liang,
Detecting community structure in complex networks via node
similarity, Physica A 389 (2010) 2849.
\bibitem{Wang2009} Y.-L. Wang, T. Zhou, J.-J. Shi, J. Wang, D.-R. He, Empirical analysis of dependence between stations in Chinese railway network, Physica A 388 (2009) 2949.
\bibitem{Zhou2007} T. Zhou, J. Ren, M. Medo, Y.-C. Zhang, Bipartite network projection and personal recommendation, Phys. Rev.E 76 (2007) 046115.

\bibitem{Katz1953}L. Katz, A new status index derived from sociometric analysis, Psychmetrika 18 (1953) 39.
\bibitem{Klein1993} D. J. Klein, M. Randic, Resistance distance, J. Math. Chem. 12
(1993) 81.
\bibitem{Fouss2007} F. Fouss, A. Pirotte, J.-M. Renders, M. Saerens, Random-walk computation of similarities between nodes of a graph
with application to collaborative recommendation, IEEE Trans. Knowl.
Data. Eng. 19 (2007) 355.
\bibitem{Brin1998} S. Brin, L. Page, The anatomy of a
large-scale hypertextual Web search engine, Comput. Netw. ISDN Syst.
30 (1998) 107.
\bibitem{Tong2006} H. Tong, C. Faloutsos, J.-Y. Pan, Fast random walk with
restart and its applications, In Proceedings of the 6th
International Conference on Data Mining, IEEE Press, Washington, DC,
USA, 2006, p. 613-622 .
\bibitem{Shang2009} M.-S. Shang, L. L\"{u}, T. Zhou, Y.-C. Zhang, Relevance is more significant than
correlation: Information filtering on sparse data, EPL 88 (2009)
68008.
\bibitem{Jeh2002} G. Jeh, J. Widom, SimRank: A measure of structural-context similarity,
In Proceedings of the ACM SIGKDD International Conference on
Knowledge Discovery and Data Mining, ACM Press, New York, 2002, p.
271-279.
\bibitem{Chebotarev1997} P. Chebotarev, E. V. Shamis, The matrix-forest theorem and measuring relations in small social groups, Automation
and Remote Control 58 (1997) 1505.
\bibitem{Fouss2006} F. Fouss, L. Yen, A. Pirotte, M. Saerens, An
experimental investigation of graph kernels on a collaborative
recommendation task, In Proceedings of the 6th International
Conference on Data Mining, IEEE Press, Washington, DC, USA, 2006, p.
863-868.

\bibitem{Lu2009} L. L\"u, C.-H. Jin, T. Zhou, Similarity index based on local paths for link prediction of complex
networks, Phys. Rev. E 80 (2009) 046122.
\bibitem{Liu2010} W. Liu, L. L\"u, Link prediction based on local
random walk, EPL 89 (2010) 58007.
\bibitem{Mantrach2010} A. Mantrach, N. van zeebroeck, P. Francq, M. Shimbo, H. Bersini, M.
Saerens, Semi-supervised classification and betweenness
computation on large, sparse, directed, networks (unpublished).

\bibitem{Clauset2008} A. Clauset, C. Moore, M. E. J. Newman, Hierarchical structure and the prediction of missing links in networks, Nature
453 (2008) 98.
\bibitem{Zhou2006} C. Zhou, L. Zemanov\'a, G. Zamora, C. C.
Hilgetag, J. Kurths, Hierarchical Organization Unveiled by
Functional Connectivity in Complex Brain Networks, Phys. Rev. Lett.
97 (2006) 238103.
\bibitem{Redner2008} S. Redner, Teasing out the missing links,
Nature 453 (2008) 47.
\bibitem{Casella2001} G. Casella, R. L. Berger, Statistical
Inference, Duxbury, Belmont, 2001.
\bibitem{Newman1999} M. E. J. Newman, G. T. Barkema, Monte Carlo
Methods in Statistical Physics, Clarendon, Oxford, 1999.
\bibitem{Krebs2002} V. Krebs, Mapping networks of
terrorist cells, Connections 24 (2002) 43.
\bibitem{Dawah1995} H. A. Dawah, B. A. Hawkins, M. F. Claridge, Structure of the parasitoid
communities of grass-feeding chalcid wasps, J. Anim. Ecol. 64 (1995)
708.
\bibitem{Huss2007} M. Huss, P. Holme, Currency and commodity
metabolites: Their identification and relation to the modularity of
metabolic networks, IET Syst. Biol. 1 (2007) 280.
\bibitem{Mossel2005} E. Mossel, E. Vigoda, Phylogenetic MCMC Are
Misleading on Mixtures of Trees, Science 309 (2005) 2207.

\bibitem{White1976} H. C. White, S. A. Boorman, R. L. Breiger,
Social structure from multiple networks I: Blockmodels of roles and
positions, Am. J. Sociol. 81 (1976) 730.
\bibitem{Holland1983} P. W. Holland, K. B. Laskey, S. Leinhardt,
Stochastic blockmodels: First steps, Social Networks 5 (1983) 109.
\bibitem{Dorelan2005} P. Dorelan, V. Batagelj, A. Ferligoj,
Generalized Blockmodeling, Cambridge University Press, Cambridge,
UK, 2005.
\bibitem{Airoldi2008} E. M. Airoldi, D. M. Blei, S. E. Fienberg, X.
P. Xing, Mixed-membership stochastic blockmodels, J. Mach. Learn.
Res. 9 (2008) 1981.
\bibitem{Guimera2007} R. Guimer\`{a}, M. Sales-Pardo, L. A. N.
Amaral, Classes of complex networks defined by role-to-role
connectivity profiles, Nat. Phys. 3 (2007) 63.
\bibitem{Newman2002} M. E. J. Newman, Assortative mixing in networks, Phys. Rev. Lett. 89 (2002)
208701.
\bibitem{Newman2003b} M. E. J. Newman, Mixing patterns in networks,
Phys. Rev. E 67 (2003) 026126.
\bibitem{Pastor-Satorras2001} R. Pastor-Satorras, A. V\'azquez, A.
Vespignani, Dynamical and Correlation Properties of the Internet,
Phys. Rev. Lett. 87 (2001) 258701.
\bibitem{Vazquez2002} A. V\'azquez, R. Pastor-Satorras, A.
Vespignani, Large-scale topological and dynamical properties of the
Internet, Phys. Rev. E 65 (2002) 066130.
\bibitem{Bayes1763} T. Bayes, An essay towards solving a Problem in
the Doctrine of Chances, Philos. Trans. R. Soc. London 53 (1763)
370.
\bibitem{Metropolis1953} M. Metropolis, A. W. Rosenbluth, A. H.
Teller, E. Teller, Equations of State Calculation by Fast Computing
Machines, J. Chem. Phys. 21 (1953) 1087.
\bibitem{Zachary1977} W. Zachary, An information flow model for conflict and
fission in small groups, J. Anthropol. Res. 33 (1977) 452.
\bibitem{Lusseau2003} D. Lusseau, \emph{et al.}, The bottlenose dolphin community of Doubtful Sound
features a large proportion of long-lasting associations, Behav.
Ecol. Sociobiol. 54 (2003) 396.
\bibitem{Guimera2005} R. Guimer\`{a}, S. Mossa, A. Turtschi, L. A.
N. Amaral, The worldwide air transportation network: Anomalous
centrality, community structure, and cities' global roles, Proc.
Natl. Acad. Sci. U.S.A. 102 (2005) 7794.
\bibitem{White1986} J. G. White, E. Southgate, J. N. Thomson, S.
Brenner, The structure of the nervous system of the nematode C.
elegans, Phil. Trans. R. Soc. B 314 (1986) 1.
\bibitem{Reed2003} J. L. Reed, T. D. Vo, C. H. Schilling, B. {\O}.
Palsson, An expanded genome-scale model of Escherichia coli K-12
(iJR904 GSM/GPR), Genome Biol. 4 (2003) R54.



\bibitem{Friedman1999} N. Friedman, L. Getoor, D. Koller, A. Pfeffer, Learning
Probabilistic relational models, In Proceedings of the 16th
International Joint Conference on Artificial Intelligence,
Stockholm, Sweden, 1999, p. 1300.
\bibitem{Heckerman2004_1} D. Heckerman, C. Meek, D. Koller, Probabilistic
Entity-Relationship Models, PRMs, and Plate Models, In Proceedings
of the 21st International Conference on Machine Learning, Banff,
Canada, 2004, p. 55.
\bibitem{Yu2006} K. Yu, W. Chu, S. Yu, V. Tresp, Z. Xu, Stochastic
Relational Models for Discriminative Link Prediction, In Proceedings
of Neural Information Precessing Systems, MIT Press, Cambridge MA,
2006, p. 1553.
\bibitem{Neville2006} J. Neville, Statistical models and analysis
techniques for learning in relational data, PhD Thesis, 2006.
\bibitem{Heckerman2004_2} D. Heckerman, C. Meek, D. Koller, Probabilistic models for
relational data, Tech. Rep. MSR-TR-2004-30, Microsoft Research,
2004.
\bibitem{Heckerman1995} D. Heckerman, D. Geiger, D. Chickering,
Learning Bayeaian Networks: the combination of knowledge and
statistical data, Machine Learning 20 (1995) 197.
\bibitem{Taskar2002} B. Taskar, P. Abbeel, D. Koller, Discriminative
probabilistic models in relational data, In Preceedings of the 18th
Conference on Uncertainty in Artificial Intelligence (UAI02),
Edmonton, Canada, 2002, p.485.
\bibitem{Taskar2003} B. Taskar, M.-F. Wong, P. Abbeel, D. Koller,
Link prediction in relational data, In Proceedings of Neural
Information Precessing Systems, MIT Press, Cambridge MA, 2004, p.
659.
\bibitem{Buntine1994} W. Buntine, Operations for learning with graphical models,
J. Artif. Intell. Res. 2 (1994) 159.
\bibitem{Spiegelhalter1998} D. Spiegelhalter, Bayesian graphical modeling: A case-study in monitoring health
outcomes, Applied Statistics 47 (1998) 115.
\bibitem{Heckerman2000} D. Heckerman, D. Chickering,
C. Meek, R. Rounthwaite, C. Kadie, Dependency networks for
inference, collaborative filtering, and data visualization, J.
Machine Learning Res. 1 (2000) 49.
\bibitem{Neville2007} J. Neville, D. Jensen, Relational dependency
networks, J. Machine Learning Res. 8 (2007) 653.
\bibitem{Casella1992} G. Casella, E. I. George, Explaining the Gibbs
sampler, The American Statistician 46(3) (1992) 167.
\bibitem{Xu2005} Z. Xu, V. Tresp, K. Yu, S. Yu, H.-P. Kriegel,
Dirichlet enhanced relational learning, In Proceedings of the 22nd
internatonal conference on machine learning, Bonn, Germany, 2005, p.
1004.
\bibitem{Yu2007} K. Yu, W. Chu, Gaussian process models for link
analysis and transfer learning, In Proceedings of Neural Information
Precessing Systems, MIT Press, Cambridge MA, 2007, p. 1657.
\bibitem{Chu2006} W. Chu, V. Sindhwani, Z. Ghahramani, S. S.
Keerthi, Relational learning with gaussian processes, In Proceedings
of Neural Information Precessing Systems, MIT Press, Cambridge MA,
2006, p. 289.

\bibitem{Madadhain2005} J. O'Madadhain, J. Hutchins, P. Smyth,
Prediction and ranking algorithms for event-based network data, In
Proceedings of SIGKDD 2005, ACM Press, New York, 2005, p. 23.
\bibitem{Shang2010} M.-S. Shang, L. L\"u, Y.-C. Zhang, T. Zhou, Empirical analysis of
web-based user-object bipartite networks, EPL 90 (2010) 48006.
\bibitem{Kunegis2010} J. Kunegis, E. W. De Luca, S. Albayrak, The
link predection problem in bipartite networks, arXiv: 1006.5367.
\bibitem{Zhou2010} T. Zhou, Z. Kuscsik, J.-G. Liu, M. Medo, J. R. Wakeling, Y.-C. Zhang, Solving the apparent diversity-accuracy dilemma of recommender systems, Proc. Natl. Acad. Sci. U.S.A. 107 (2010)
4511.
\bibitem{Zeng2010} W. Zeng, M.-S. Shang, Q.-M. Zhang, L. L\"u, T. Zhou,
Can dissimilar users contribute to accuracy and diversity of
personalized recommendation, Int. J. Mod. Phys. C (to be published).
\bibitem{Schafer2001} J. Schafer, J. Konstan, J. Riedl, E-commerce recommendation applications, Data Min.
Knowl. Discovery 5 (2001) 115.
\bibitem{Huang2006} Z. Huang, D. D. Zeng, A link prediction approach to anomalous email
detection, In Proceedings of 2006 IEEE International Conference on
Systems, Man, and Cybernetics, Taipei, Taiwan, 2006, p. 1131.
\bibitem{Gallagher2008} B. Gallagher, H. Tong, T. Eliassi-Rad, C.
Faloutsos, Using ghost edges for classification in sparsely labeled
networks, In Proceedings of the ACM SIGKDD International Conference
on Knowledge Discovery and Data Mining, ACM Press, New York, 2008,
p. 256.
\bibitem{Dasgupta2008} K. Dasgupta, R. Singh, B. Viswanathan,
D. Chakraborty, S. Mukherjea, A. A. Nanavati, A. Joshi, Social ties
and their relevance to churn in mobile telecom networks, In
Proceedings of the 11th international conference on Extending
database technology: Advances in database technology, ACM Press, New
York, 2008, p. 668.
\bibitem{Newman2004} M. E. J. Newman, M. Girvan, Finding and
evaluating community structure in networks, Phys. Rev. E 69 (2004)
026113.
\bibitem{Guimera2002} R. Guimer\`a, A. D\'iaz-Guilera, F.
Vega-Redondo, A. Cabrales, A. Arenas, Optimal network topologies for
local search with congestion, Phys. Rev. Lett. 89 (2002) 248701.
\bibitem{Yan2006} G. Yan, T. Zhou, B. Hu. Z.-Q. Fu, B.-H. Wang, Efficient routing on complex
networks, Phys. Rev. E 73 (2006) 046108.
\bibitem{Barahona2002} M. Barahona, L. M. Pecora, Synchronization in
Small-World Systems, Phys. Rev. Lett. 89 (2002) 054101.
\bibitem{Arenas2008} A. Arenas, A. D\'iza-Guilera, J. Kurths, Y. Moreno,
C. Zhou, Phys. Rep. 469 (2008) 93.
\bibitem{Pastor-Satorras2003} R. Pastor-Satorras, A. Vespignani,
Epidemics and immunization in scale-free networks, in S. Bornholdt,
H. G. Schuster (eds.) Handbook of Graphs and Networks, Wiley-VCH,
Berlin, 2003.
\bibitem{Zhou2006b} T. Zhou, Z.-Q. Fu, B.-H. Wang, Epidemic dynamics on complex
networks, Prog. Nat. Sci. 16 (2006) 452.
\bibitem{Caldarelli2002} G. Caldarelli, A. Capocci, P. De Los Rios, M. A. Mu\~noz, Scale-free networks from varying vertex intrinsic fitness, Phys. Rev. Lett. 89 (2002) 258702.
\bibitem{Valverde2002} S. Valverde, R. F. Cancho, R. V. Sol\'e, Scale-free networks from optimal
design, Europhys. Lett. 60 (2002) 512.
\bibitem{Baiesi2003} M. Baiesi, S. S. Manna, Scale-free networks from a Hamiltonian dynamics, Phys. Rev. E 68 (2003) 047103.
\bibitem{Kim2005} B. J. Kim, A. Trusina, P. Minnhagen, K. Sneppen, Self organized scale-free networks from merging and regeneration, Eur. Phys. J. B 43 (2005) 369.
\bibitem{Perotti2009} J. I. Perotti, O. V. Billoni, F. A. Tamarit, D. R. Chialvo, S. A. Cannas, Emergent self-organized complex network topology out of stability constraints, Phys. Rev. Lett. 103 (2009) 108701.
\bibitem{Liu2008} H.-K. Liu, T. Zhou, Review on the studies of
airline networks, Prog. Nat. Sci. 18 (2008) 601.
\bibitem{Cui2010} A.-X. Cui, Y. Fu, M.-S. Shang, D.-B. Chen, T.
Zhou, Emergence of local structures in complex network:common
neighborhood drives the network evolution, Acta Physica Sinica (to
be published).
\bibitem{Xiao2007} W.-K. Xiao, J. Ren, F. Qi, Z.-W. Song, M.-X. Zhu,
H.-F. Yang, H.-Y. Jin, B.-H. Wang, T. Zhou, Emprical study on
clique-degree distribution of networks, Phys. Rev. E 76 (2007)
037102.
\bibitem{Lambiotte2008}R. Lambiotte, V. D. Blondel, C. de Kerchove, E. Huens, C. Prieur, Z. Smoreda, P. Van Dooren,
Geographical dispersal of mobile communication networks, Physica A
387 (2008) 5317.
\bibitem{Jung2008} W.-S. Jung, F. Wang, H.E. Stanley, Gravity model in the korean highway, EPL 81 (2008)
48005.
\bibitem{Kaluza2010} P. Kaluza, A. Koelzsch, M. T. Gastner, B. Blasius, The complex network of global cargo ship movements, J. R. Soc. Interface 7 (2010) 1093.

\bibitem{Bianconi2009} G. Bianconi, P. Pin, M. Marsili, Assessing the relevance of node features for network
structure, Proc. Natl. Acad. Sci. U.S.A. 106 (2009) 11433.
\bibitem{Liu2009} H.-K. Liu, X.-L. Zhang, L. Cao, B.-H. Wang, T.
Zhou, Analysis on the connecting mechanism of Chinese city airline
network, Sci. China Ser. G 39 (2009) 935.
\bibitem{Granger1969} C. W. J. Granger, Investigating causal relations by econometric models and cross-spectral methods, Econometrica 37 (1969) 424.
\bibitem{Liu2010b} H.-K. Liu, X.-L. Zhang, T. Zhou, Structure and External Factors of
Chinese City Airline Network, Physics Procedia 3 (2010) 1781.
\bibitem{ZhangQM2010} Q.-M Zhang, M.-S. Shang, L. L\"u,
Similarity-based classification in partially labeled networks, Int.
J. Mod. Phys. C 21 (2010) 813.


\bibitem{Alon2007} U. Alon, Network motifs: theory and experimental
approaches, Nat. Rev. Gene. 8 (2007) 450.
\bibitem{Mantrach2010b} A. Mantrach, L. Yen, J. Callut, K.
Fran\c{c}oisse, M. Shimbo, M. Saerens, The Sum-over-Paths
Covariance Kernel: A Novel Covariance Measure between Nodes of a
Directed Graph, IEEE Trans. Pattern Analysis and Machine
Intelligence 32 (2010) 1112.
\bibitem{Murata2007} T. Murata, S. Moriyasu, Link prediction of
social networks based on weighted proximity measure, In Proceedings
of the IEEE/WIC/ACM International Conference on Web Intelligence,
ACM Press, New York, 2007.
\bibitem{Lu2010} L L\"u, T. Zhou, Link prediction in weighted
networks: The role of weak ties, EPL 89 (2010) 18001.
\bibitem{Yin2002} H. Yin, S. C. Wong, J. Xu, C. K. Wong, Urban traffic flow prediction using a fuzzy-neural
approach, Transportation Res. C 10 (2002) 85.
\bibitem{Kunegis2009} J. Kunegis, A. Lommatzsch, C. Bauckhage, The
Slashdot Zoo: Mining a social network with negative edges, In
Proceedings of WWW'2009, ACM Press, New York, 2009.
\bibitem{Guha2004} R. V. Guha, R. Kumar, P. Raghavan, A. Tomkins,
Propagation of trust and distrust, In Proceedings of WWW'2004, ACM
Press, New York, 2004.
\bibitem{Leskovec2010} J. Leskovec, D. Huttenlocher, J. Kleinberg,
Predicting Positive and Negative Links in Online Social Networks, In
Proceedings of WWW'2010, ACM Press, New York, 2010.
\bibitem{Traag2009} V. A. Traag, J. Bruggeman, Community detection
in networks with positive and negative links, Phys. Rev. E 80 (2009)
036115.
\bibitem{Marvel2009} S. A. Marvel, S. H. Strogatz, J. M. Kleinberg,
Energy Landscape of Social Balance, Phys. Rev. Lett. 103 (2009)
198701.
\bibitem{Szell2010} M. Szell, R. Lambiotte, S. Thurner,
Multirelational organization of large-scale social networks in an
online world, Proc. Natl. Acad. Sci. U.S.A. 107 (2010) 13636.
\bibitem{Zhang2010} Z.-K. Zhang, T. Zhou, Y.-C. Zhang, Personalized
recommendation via integrated diffusion on user-item-tag tripartite
graphs, Physica A 389 (2010) 179.
\bibitem{Burke2002} R. Burke, Hybrid recommender systems: Survey and
experiments, User Model. User-Adap. Interact. 12 (2002) 331.
\bibitem{Polikar2006} R. Polikar, Ensemble based systems in decision
making, IEEE Circuits and Systems Magazine 6(3) (2006) 21.

\bibitem{Leroy2010} V. Leroy, B. B. Cambazoglu, F. Bonchi, Cold
start link prediction, In Proceedings of the 16th ACM SIGKDD
International Conference on Knowledge Discovery and Data Mining, ACM
Press, New York, 2010, p. 393.
\bibitem{Zheleva2008} E. Zheleva, L. Getoor, J. Golbeck, Ugur Kuter,
Using friendship ties and family circles for link prediction, In
Proceedings of the 2nd Workshop on Social Network Mining and
Analysis, ACM Press, New York, 2008.
\bibitem{Cao2010} B. Cao, N. N. Liu, Q. Yang, Transfer learning for
collective link prediction in multiple heterogenous domains, In
Proceedings of the 27th International Conference on Machine
Learning, Haifa, Israel, 2010.
\bibitem{Huang2009} Z. Huang, D. K. J. Lin, The time-series link
prediction problem with applications in communication surveillance,
INFORMS J. Comput. 21 (2009) 286.
\bibitem{Tylenda2009} T. Tylenda, R. Angelova, S. Bedathur, Towards
time-aware link prediction in evolving social networks, In
Proceedings of the 3rd Workshop on Social Network Mining and
Analysis, ACM Press, New York, 2009.



\end{thebibliography}
\end{document}